\documentclass[aps,prx,twocolumn,superscriptaddress,longbibliography,floatfix]{revtex4-2}

\usepackage{hyperref}
\hypersetup{
	colorlinks=true,
	linkcolor=blue,
	filecolor=blue,
	citecolor = black,      
	urlcolor=cyan,
}
\usepackage{ulem}
\usepackage{amsmath}
\usepackage{amsfonts}
\usepackage{graphicx}
\usepackage{siunitx}

\usepackage{xargs}
\usepackage[pdftex,dvipsnames]{xcolor}
\usepackage{color}

\usepackage{empheq}
\usepackage[shortlabels]{enumitem}

\usepackage{xspace}

\newcommand{\magn}[1]{\left|#1\right|}

\newcommand{\bra}[1]{\left\langle#1\right|}
\newcommand{\ket}[1]{\left|#1\right\rangle}

\newcommand{\figref}[1]{Fig.~\ref{#1}}

\newcommand{\tr}[1]{\mathrm{Tr}\left[ #1 \right]}
\newcommand{\trover}[2]{\mathrm{Tr}_{#1}\left[ #2 \right]}

\newcommand{\rcode}{r_{\mathrm{code}}}
\newcommand{\dcode}{d_{\mathrm{code}}}

\newcommand{\pwrk}{$\mathrm{PWR}2_k$\xspace}
\newcommand{\pwr}{$\mathrm{PWR}2$\xspace}
\newcommand{\pchaar}{p_{c,\mathrm{Haar}}}
\newcommand{\cz}{\mathrm{CZ}}
\newcommand{\stabs}{\mathcal{S}}
\newcommand{\centr}{\mathcal{C}(\mathcal{S})}
\newcommand{\proj}{\mathrm{proj}}

\newcommand{\projab}{\mathrm{proj}_{\overline{A}}}
\newcommand{\id}{\mathbb{I}}

\newcommand{\sub}{\mathrm{sub}}

\begin{document}
\date{\today}
\title{Measurement-Induced Phase Transitions in Sparse Nonlocal Scramblers}

\author{Tomohiro Hashizume}
\affiliation{Department of Physics and SUPA, University of Strathclyde, Glasgow G4 0NG, United Kingdom}
\author{Gregory Bentsen}
\affiliation{Martin A. Fisher School of Physics, Brandeis University, Waltham, Massachusetts 02465, USA}
\author{Andrew J.~Daley}
\affiliation{Department of Physics and SUPA, University of Strathclyde, Glasgow G4 0NG, United Kingdom}

\begin{abstract}
    Measurement-induced phase transitions arise due to a competition between the scrambling of quantum information in a many-body system and local measurements. In this work we investigate these transitions in different classes of fast scramblers, systems that scramble quantum information as quickly as is conjectured to be possible -- on a timescale proportional to the logarithm of the system size. In particular, we consider sets of deterministic sparse couplings that naturally interpolate between local circuits that slowly scramble information and highly nonlocal circuits that achieve the fast-scrambling limit. We find that circuits featuring sparse nonlocal interactions are able to withstand substantially higher rates of local measurement than circuits with only local interactions, even at comparable gate depths. We also study the quantum error-correcting codes that support the volume-law entangled phase and find that our maximally nonlocal circuits yield codes with nearly extensive contiguous code distance.
    Use of these sparse, deterministic circuits opens pathways towards the design of noise-resilient quantum circuits and error correcting codes in current and future quantum devices with minimum gate numbers.  
\end{abstract}

\maketitle

\section{Introduction}
\label{sec:intro}

Measurement-induced phase transitions (MIPT) \cite{li2018quantum,skinner2019measurement,chan2019unitary,aharonov2000quantum} are transitions driven by a competition between scrambling dynamics, which tends to generate many-body entanglement across all degrees of freedom of a quantum many-body system, and local measurements, which tend to destroy this entanglement. Measurement-driven transitions of this type have garnered an extraordinary amount of theoretical interest recently \cite{li2019measurement,szyniszewski2019entanglement,bao2020theory,jian2020measurement,fan2020self,li2020conformal,choi2020quantum,zabalo2020critical,tang2020measurement,turkeshi2020measurement,zhang2020nonuniversal,goto2020measurement,li2020statistical,botzung2021engineered,mendoza2021self,mendoza2021self,li2021entanglement,fidkowski2021dynamical}, and have been studied in a wide range of physical systems, including in random quantum circuits, free fermions \cite{cao2018entanglement,chen2020emergent,alberton2021entanglement}, and 
non-Hermitian Brownian models \cite{bentsen2021measurement}. They have also been extended to systems with long-range interactions \cite{iaconis2020measurement,block2021measurement,zhang2021universal,muller2021measurement}, including models with all-to-all coupling \cite{nahum2020measurement,vijay2020measurement}, and non-Hermitian Sachdev-Ye-Kitaev (SYK) models \cite{liu2020non,zhang2021syk,jian2021syk}. MIPTs have also been related to purification transitions \cite{gullans2020dynamical}, where entanglement of the system undergoing MIPT with an ancilla can be used to detect the transition \cite{gullans2020scalable}, as recently demonstrated in experiments with trapped ions \cite{noel2021observation}. Importantly, these transition have also been related to the properties of quantum error-correcting codes (QECC) \cite{choi2020quantum,gullans2020dynamical}. These codes are dynamically generated during the evolution and are responsible for protecting the extensive many-body entanglement present in the system despite repeated local measurements. 

The durability of this many-body entangled phase, despite the presence of repeated projective measurements, is particularly compelling from the perspective of experiments with modern quantum technologies. In this context, the goal is often to generate interesting or useful entangled quantum states, while always contending with noise and dissipation which inevitably degrade this entanglement \cite{aharonov2000quantum}. In this context, it is therefore of considerable value to identify mechanisms for building up robust patterns of entanglement as quickly and as efficiently as possible before noise has a chance to degrade the computation or simulation being carried out. Particularly efficient quantum systems that build up many-body entanglement as rapidly as possible are known as \emph{fast scramblers} \cite{sekino2008fast,lashkari2013towards}.
Fast scrambling was first discussed in the context of information spreading in black holes, and refers to systems that can scramble quantum information on timescales $t_* \sim \log N$ that grow only logarithmically with the system size $N$ \cite{page1993average,hayden2007black,sekino2008fast,lashkari2013towards,hosur2016chaos}. Random all-to-all circuits \cite{piroli2020random}, and other disordered all-to-all coupled models such as the SYK model \cite{ye1993solvable,kitaev2015} are good examples of fast scramblers, but recently, a number of deterministic and experimentally feasible systems exhibiting fast scrambling have also been explored \cite{bentsen2019treelike,belyansky2020minimal,li2020fast,hashizume2021deterministic}. These show similar dynamics, but often on sparse nonlocal coupling graphs. In this context, we are led to the natural questions: how do these deterministic fast scrambling circuits perform when subjected to random measurements? Does fast scrambling dynamics uniformly enhance the stability of the mixed phase as a general principle, and can this lead to improved quantum error-correcting code properties?

In this work we explore these connections by studying how the strength of scrambling affects the properties of the MIPT mixed phase by considering sparsely-coupled quantum circuits with tunable nonlocal interactions. By adjusting the range of interactions in these sparse circuits, we can tune from local spreading of information to genuine fast scrambling. 
Our analysis demonstrates that sparse nonlocal interactions can significantly improve a quantum system's robustness to local measurements. We show that 
only a few additional layers of nonlocal gates substantially increase the critical measurement rate $p_c$. We further demonstrate that highly nonlocal interactions improve the properties of the quantum error-correcting codes that stabilize the mixed phase.

\begin{figure}[h!]
    \centering
    \includegraphics[width=\columnwidth]{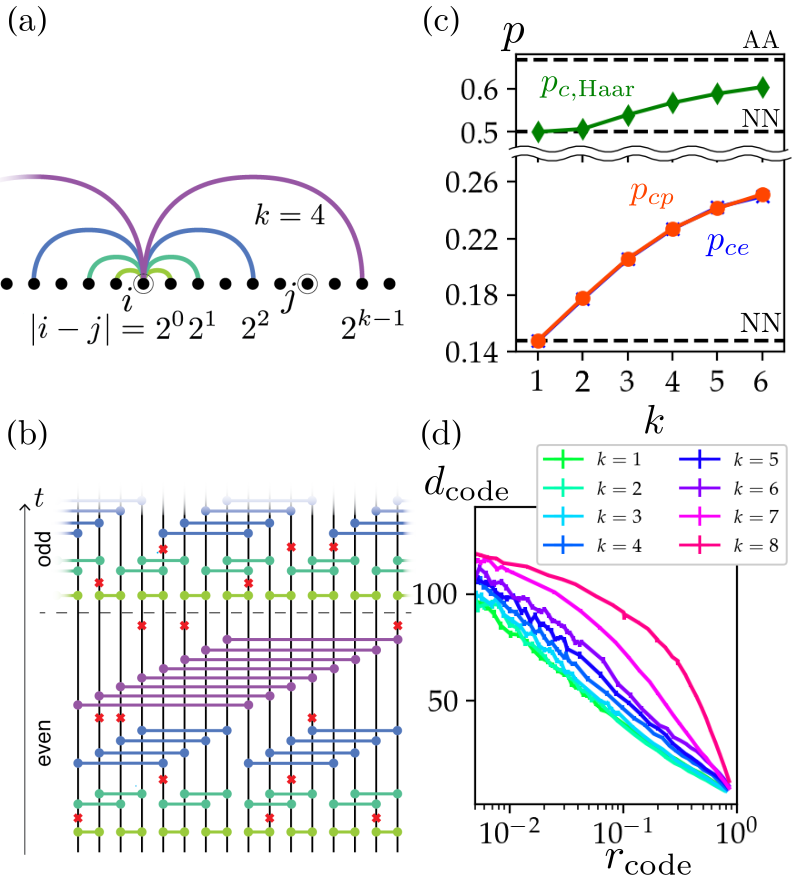}
    \caption{Sparse fast scrambling \pwrk circuits with tunable nonlocality $k$ subjected to random projective measurements.
    (a,b) Two-qubit entangling gates $Q_{ij}$ (green, turquoise, blue, purple) are applied between qubits $i,j$ if and only if they are separated by an integer power of 2, $\magn{i-j} = 2^{m-1}$ for $m = 1,\ldots,k$. These gates are applied in consecutive interaction layers at times $t = 1,\ldots,T$ in a nonlocal bricklayer pattern arranged into alternating even and odd blocks up to a total depth $T = 8N$.
    Projective single-qubit measurements (red crosses) occur with probability $p$ between each interaction layer. The degree $k$ of the resulting interaction graph (b) controls the nonlocality of interactions where the longest-range interactions occur between qubits initially separated by $d_{\mathrm{max}} = \magn{i-j} = 2^{k-1}$. (c) The critical measurement rates for the purification transition $p_{cp}$ (red) and entanglement transition $p_{ce}$ (blue) increase significantly with $k$, interpolating between the critical points found for random nearest neighbor (NN) models for $k = 1$ and random all-to-all (AA) models for larger $k$. Haar-random circuits with identical coupling patterns show a similar improvement in robustness to measurement as $k$ increases (green). (d) In the mixed phase for $N = 256$ qubits, the maximum code distance $\dcode$ improves with $k$ at fixed code rate $\rcode$. Error bars are shown or are smaller than datapoints; lines are guides to the eye.}
    \label{fig:SparseMIPTFig1}
\end{figure}

As specific examples, we choose sparse nonlocal circuits that can be realised in the laboratory by combining nearest-neighbor Rydberg interactions with nonlocal shuffling operations \cite{hashizume2021deterministic}. The strength of scrambling in these circuits is governed by a nonlocality parameter $k$. We refer to this family of circuits as `power-of-2' (\pwrk) models because they feature sparse long-range couplings only between qubits separated by an integer power of 2, up to a maximum distance $d_{\mathrm{max}} = 2^{k-1}$. In experimental implementations with Rydberg atoms the nonlocality parameter $k$ is controlled by the number of shuffle operations applied, thereby providing a natural tunable parameter controlling the strength of scrambling in these circuits.
In particular, the case $k=1$ corresponds to a nearest neighbour (NN) circuit with slow scrambling, while $k \sim \log_2 N$ yields highly nonlocal circuits which rapidly scramble quantum information. In the following we study how these circuits perform when subjected to random projective measurements with a view toward understanding how nonlocal interactions can improve the stability of the MIPT mixed phase.

The rest of this paper is organised as follows: In Section \ref{sec:models} we introduce the deterministic \pwrk circuit models that we study in this work. In Section \ref{sec:percolation} we study measurement-induced transitions in \pwrk circuits featuring Haar-random gates, which allows the problem to be mapped to a classical percolation problem. In Section \ref{sec:entanglement} we study the entanglement transition numerically in \pwrk circuits featuring Clifford gates and characterize the transition in terms of the tripartite mutual information $I(A:B:C)$ between three consecutive regions of the output qubits. In Section \ref{sec:purification} we study the purification transition numerically for the same Clifford circuits as characterized by the purification time $\tau$ of a single qubit. In Section \ref{sec:qecc} we numerically study the properties of the dynamically-generated error-correcting code in the mixed phase, and demonstrate that nonlocal interactions allow for improved code distance at fixed code rate. In Section \ref{sec:nonlocalaa} we consider the `complete' \pwrk circuit with $k = \log_2 N$ and compare its behavior to the random all-to-all model. Finally, we discuss outstanding issues and prospects for future work in Section \ref{sec:conclusion}.

\section{Models}
\label{sec:models}

The models we study here feature sparse nonlocal interactions in which pairs of spins residing at sites $i,j = 1,\ldots,N$ of a 1d lattice are coupled if and only if the distance between them is an integer power of 2: $\magn{i-j} = 2^{m-1}$ for $m = 1,\ldots,k$ as shown in \figref{fig:SparseMIPTFig1}a. We refer to these interactions as `nonlocal' in the sense that qubits $i,j$ in the chain may be coupled even when they are far away from one another, but we emphasize that all couplings are `2-local' in the quantum information sense because each gate acts on only 2 qubits at a time. Sparse nonlocal interactions of this type have been shown to generate fast scrambling dynamics  \cite{bentsen2019treelike}, and are inspired by the nonlocal coupling patterns that can be engineered in single-mode cavities \cite{periwal2021programmable} or in Rydberg arrays with the aid of tweezer-assisted shuffling operations \cite{hashizume2021deterministic}. 
We expect the fast scrambling dynamics in these nonlocal circuits to rapidly generate many-body entanglement that is increasingly robust to the destructive effects of local measurements relative to circuits featuring only local interactions.

In the following we demonstrate these expectations explicitly by studying sparse nonlocal Floquet circuits consisting of consecutive layers of pairwise nonlocal two-qubit gates $Q_{ij} = Q_{ji}$. The gates are arranged in a nonlocal bricklayer pattern, as illustrated in \figref{fig:SparseMIPTFig1}b, where each timestep $t = 1,\ldots,T$ represents a single interaction layer consisting of exactly $N/2$ gates. These interaction layers are stacked into an alternating sequence of even and odd interaction blocks. During the \emph{even} block $t = m = 1,\ldots,k$ we place gates $Q_{ij}$ between qubits 
$i < j$ if and only if $\magn{i-j} = 2^{m-1}$ and $\textrm{mod}(\lfloor i / 2^{m-1} \rfloor,2) = 0$. During the subsequent \emph{odd} block $t = k + m =  k+1,\ldots,2k$ we place gates according to the same rules but with the odd-bricklayer condition $\textrm{mod}(\lfloor i / 2^{m-1} \rfloor,2) = 1$. Finally, we apply a layer of single-qubit phase gates $P_i$ to all qubits and repeat the entire sequence until a final time $t = T$. Unless otherwise specified, we consider circuits with total depth $T = 8N$ for Clifford gates and $T = N/2$ for Haar-random gates. This alternating sequence of even and odd nonlocal interaction blocks ensures that quantum information rapidly spreads through the system via nonlocal interactions as rapidly as possible within a single block, but is not siloed into one of the hierarchical branches of the nonlocal circuit that occurs if, for example, the circuit consists only of even blocks. Unless otherwise stated, the models which appear in this paper always assume periodic boundary conditions.

Following each interaction layer $t$ we randomly apply projective measurements in the Pauli-$z$ basis to individual qubits with probability $p$ as illustrated by the red crosses in Fig. \ref{fig:SparseMIPTFig1}b. These projective measurements tend to destroy the long-range entanglement built up by the previous interaction layers of the \pwrk circuit.
The competition between these two effects leads to a measurement-induced phase transition in the entanglement structure of the output state \cite{bao2020theory,gullans2020dynamical,nahum2020measurement}. In this paper, we are particularly interested in how this transition changes as we add more nonlocal interactions.

In the \pwrk family, this degree of nonlocality is controlled by the parameter $k \leq \log_2 N$, which determines the distance of the longest-range interactions $d_{\mathrm{max}} = \magn{i-j}_{\mathrm{max}} = 2^{k-1}$. Thus, \pwrk circuits with $k \sim O(1)$ consist only of short-range couplings which restrict the spread of quantum information \cite{lieb1972finite,hastings2006spectral} and therefore generate only slow information scrambling. By contrast, circuits with $k \sim \log_2 N$ consist of highly nonlocal interactions that span the entire system and are known to rapidly scramble quantum information \cite{bentsen2019fast,bentsen2019treelike,hashizume2021deterministic}. The nonlocality parameter $k$ therefore provides a convenient means by which to interpolate between the local and nonlocal limits of the \pwrk family.

These two extreme cases can be compared to two prototypical scrambling circuits known to exhibit measurement-induced transitions and that have been studied extensively in existing literature. For local interactions $k \sim O(1)$ we compare to a random 1+1D nearest-neighbor (NN) circuit \cite{gullans2020dynamical} in which the scrambling of quantum information is highly restricted by local Lieb-Robinson bounds \cite{lieb1972finite,hastings2006spectral,bentsen2019fast}. As we increase the parameter $k$, additional long-range interactions are switched on, leading to a quasi-one-dimensional system of $k$-dimensional hypercubes which locally scramble quantum information (see Appendix \ref{app:pwrkdimensionality}). For maximally nonlocal interactions $k \sim \log_2 N$ the entire system is a single $k$-dimensional hypercube (along with some additional couplings) which we expect to achieve fast scrambling dynamics. As such, we compare the maximally nonlocal case to a random all-to-all (AA) circuit in which qubits interact pairwise at random without regard to their spatial location \cite{gullans2020dynamical,nahum2020measurement}. These maximally nonlocal AA models were first proposed by Sekino and Susskind \cite{sekino2008fast} as prototypical examples of
\emph{fast scramblers}, systems which scramble quantum information at the fastest rate allowed by quantum mechanics.

In the remainder of this paper, we study the entanglement properties of these monitored \pwrk circuits as a function of measurement rate $p$ and nonlocality $k$ using a combination of analytical and numerical methods. We consider two types of gates $Q_{ij}$ in this paper. In Sec. \ref{sec:percolation} we take $Q_{ij}$ to be Haar-random unitary gates acting between pairs of qudits with Hilbert space dimension $q$, which maps to a classical bond percolation problem on a degree-4 network in the limit $q \rightarrow \infty$. In Secs. \ref{sec:entanglement} - \ref{sec:qecc} we take $Q_{ij} = \cz_{ij} H_i H_j$, where $\cz_{ij} = \cz_{ji}$ is the controlled-$Z$ gate on qubits $i,j$ and $H_i,P_i$ are the Hadamard and Phase gates on qubit $i$, respectively. In this case the entire circuit is composed of Clifford gates and can therefore be efficiently simulated on a classical computer \cite{gottesman1998heisenberg,aaronson2004improved}.

For the Clifford-only circuits, we detect the measurement-induced phase transition using a standard set of diagnostic tools. In Section \ref{sec:entanglement} we characterize the transition using the tripartite mutual information $I(A:B:C)$ between three consecutive regions $A,B,C$ at the final time $t = T$ \cite{gullans2020dynamical}. In Section \ref{sec:purification} we study a closely related diagnostic, the single-qubit purification time $\tau$, which characterizes how long a reference qubit will stay entangled with the system \cite{gullans2020scalable,block2021measurement}. One can view the circuit $V$ as a tensor network that is being torn apart by repeated projective measurements, where the mutual information $I(A:B:C)$ captures the connectivity of the network at the final time $t = T$ while the single-qubit purification time $\tau$ captures the connectivity of the network between the initial and final times $t = 0,T$. In each of these cases we find that the critical measurement rate $p_c$ significantly improves with $k$, indicating that nonlocal interactions substantially improve a deterministic quantum circuit's ability to withstand local measurements by leveraging nonlocal scrambling interactions. Finally, in Section \ref{sec:qecc} we study the quantum error-correcting code properties that support the volume-law phase.

\section{Percolation Transition in Haar-Random \pwrk Circuits}
\label{sec:percolation}

\begin{figure}
    \centering
    \includegraphics[width=0.95\columnwidth]{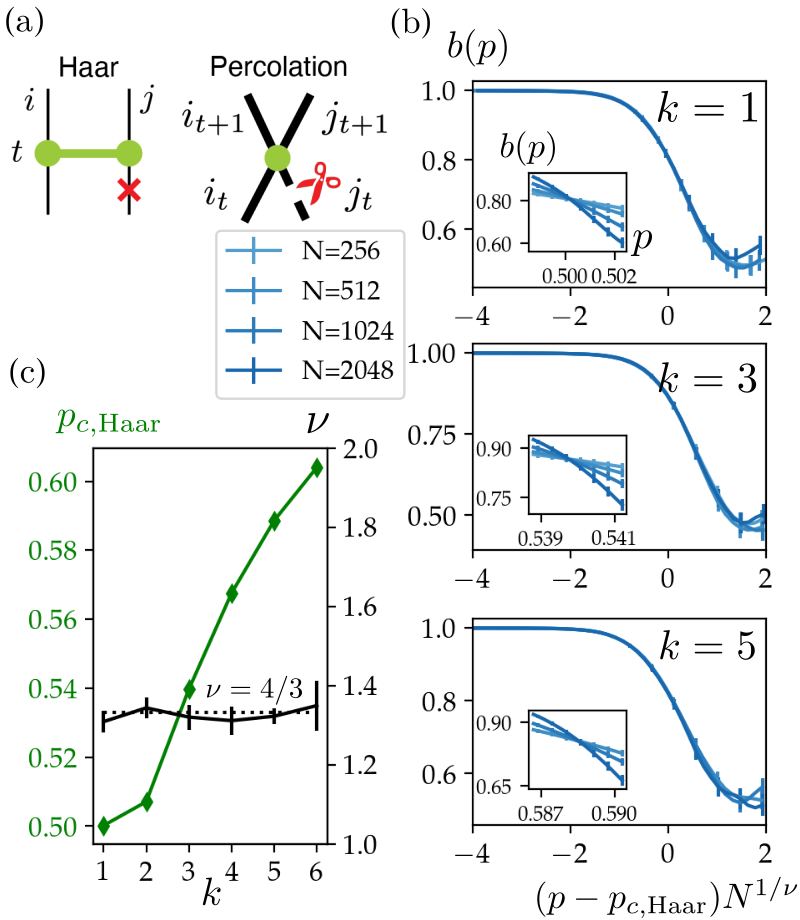}
    \caption{
        Measurement-induced transitions in the \pwrk circuit family with Haar-random gates. (a) Nonlocal circuits featuring Haar-random gates can be mapped to a classical percolation network, where gates in the original circuit (green) correspond to sites in the percolation network, and projective measurements in the Haar-random circuit (red crosses) correspond to cut bonds in the percolation network. 
        (b) The Binder cumulant $b(p)$ of the maximum cluster size in the classical percolation network for $k = 1,3,5$ (top to bottom) and system sizes $N = 256,\ldots,2048$ (light to dark blue). The critical point $\pchaar$ is extracted from the crossing point of $b(p)$ across finite-size systems (insets). 
        We observe good scaling collapse with critical exponent $\nu=4/3$ expected for a 1+1D NN circuit even for large nonlocality $k$.
        (c) The percolation critical point $\pchaar$ (green) increases with $k$; the critical exponent $\nu$ (black) is nearly constant as $k$ varies, indicating that these nonlocal circuits near criticality likely belong to the same universality class for all $k < \log_2 N$. Error bars are shown or are smaller than markers; lines are guides to the eye.
        \label{fig:Percolation_Transition_Full}
    }
\end{figure}

To demonstrate that nonlocal sparse interactions can improve a system's robustness to local measurements, we first study \pwrk circuits consisting of Haar-random nonlocal gates $Q_{ij}$ acting between pairs of qudits with local Hilbert space dimension $q$.
In the limit $q \rightarrow \infty$, the circuit and the randomly applied projective measurements can be mapped to a bond percolation problem
\cite{skinner2019measurement,gullans2020dynamical,nahum2020measurement}.
We illustrate this mapping in \figref{fig:Percolation_Transition_Full}a, where each two-qubit gate $Q_{ij}$ (green) corresponds to a vertex in the percolation network while the qubit worldlines (black) correspond to four bonds connected to this vertex. Projective measurements (red) correspond to cutting these bonds.
We discuss this mapping more explicitly in Appendix \ref{app:haarpercolation}.
Applying this mapping to the entire Floquet circuit illustrated in \figref{fig:SparseMIPTFig1}b yields a nonlocal bond percolation network 
whose bonds are cut with probability $p$.

For sufficiently small $p$ this nonlocal network percolates, in the sense that the network contains at least one connected component that extends 
from the input qubit bonds at $t = 0$ to the output bonds at $t = T$.
In the thermodynamic limit $N \rightarrow \infty$ and for extensive time $T = N/2$, this corresponds to a network with at least one connected component of infinite extent.
The percolation critical point, $\pchaar$, is defined as the maximum measurement rate $p$ for which the lattice percolates in the thermodynamic limit. For $p < \pchaar$, successful percolation implies that the output qubits at $t = T$ retain partial information about the input state at $t = 0$.
By contrast, for $p > \pchaar$, failure to percolate indicates that any information embedded in the initial state at $t = 0$ is lost due to the proliferation of projective measurements in the circuit.

Classical bond-percolation transitions of this type are well-studied in the literature \cite{staufferIntroductionPercolationTheory1994}, including on a wide variety of local and nonlocal networks.
In the extreme local case of nearest-neighbor interactions (i.e. $k = 1$ in the \pwrk family), 
the percolation mapping yields a bond percolation problem on a two dimensional square lattice, whose critical properties are known analytically \cite{kesten1980critical,staufferIntroductionPercolationTheory1994}.
In this limit, one can naturally define a correlation length $\xi$ which captures the typical size of connected components in the network. Near the critical point $\pchaar^{\mathrm{NN}}=1/2$ the correlation length diverges algebraically like $\xi\sim |p-\pchaar|^{-\nu}$, where $\nu=4/3$ is the critical exponent \cite{kesten1980critical,staufferIntroductionPercolationTheory1994,
skinner2019measurement}. 

The classical bond-percolation transition has also been studied in the opposite nonlocal limit where interactions become arbitrarily long-range. 
The percolation critical point of the random AA model was studied early on by Gullans and Huse \cite{gullans2020dynamical} 
and more extensively by Nahum \textit{et al.} \cite{nahum2020measurement}.
The critical point of the AA network is known to be at $\pchaar^{\mathrm{AA}}=2/3$.
Such a high critical point comes from the system's locally tree-like structure in the limit of large system size,
where the probability of the network having a local loop vanishes as $1/N$ \cite{nahum2020measurement}.

Here we interpolate between these two extreme limits using the nonlocality parameter $k$ in the Haar-random \pwrk circuit family, where $k = 1$ corresponds to the NN case and $k \sim \log_2 N$ approaches the AA case. 
In order to find the critical point $\pchaar$ for each value of $k$, we numerically simulate bond percolation on the corresponding network using the Newman-Ziff Algorithm \cite{newman2001fast}. The critical point $\pchaar$ and the critical exponent $\nu$ of the correlation length can be identified by calculating the binder cumulant \cite{binder1981finite,melchert2016site}, 
\begin{align}
    b(p) = \frac{1}{2}\left( 3 - \frac{\left\langle C_{\max}^4(p) \right\rangle}{\left\langle C_{\max}^2(p) \right\rangle^2}\right)
\end{align}
where $C_{\max}$ is the maximum cluster size in the nonlocal percolation network and $\langle \dots \rangle$ denotes the averaging over different configurations of randomly-cut bonds. As mentioned above, for finite $k < \log_2 N$ the system is a quasi-one-dimensional system of $k$-dimensional hypercubes arranged on a line (see Appendix \ref{app:pwrkdimensionality}), so we expect the critical properties to be governed by the usual 1+1D critical exponent $\nu$.
Near the critical point, we expect the Binder cumulant to obey the scaling law \cite{binder1981finite,melchert2016site,privman1984universal}
\begin{align}
    b(p)=f\left((p-\pchaar) N^{1/\nu} \right), \label{eq:binderscaling}
\end{align}
which is governed by the same critical exponent 
$\nu$ that controls the divergence of the correlation length $\xi$ near the critical point.

We plot the results of these numerical simulations in \figref{fig:Percolation_Transition_Full}b for $k = 1,3,5$ across system sizes $N = 64,\ldots,1024$ (light to dark blue). To extract the critical point $\pchaar$ we plot the Binder cumulant $b(p)$ (insets of \figref{fig:Percolation_Transition_Full}b) and locate the crossing point as a function of system size $N$. At each fixed $k$ in this analysis, we only consider the crossing point for sufficiently large system sizes $N > 2^k$, such that the longest-range interactions $d_{\mathrm{max}} = 2^{k-1}$ are never extensive. We then fit the scaling form \eqref{eq:binderscaling} to the data near the critical point and use this to extract an estimate of the critical exponent $\nu$ for each value of $k$.
Our resulting estimates for the critical point and critical exponent can be used to collapse the Binder cumulant $b(p)$ near the critical point to a single universal curve \eqref{eq:binderscaling} as shown in the main panels of \figref{fig:Percolation_Transition_Full}b.

We plot the resulting critical points $\pchaar$ and critical exponents $\nu$ as a function of $k$ in \figref{fig:Percolation_Transition_Full}c. 
The critical points clearly increase with nonlocality $k$; for $k \geq 5$, the critical point is closer to the AA limit than the NN limit.
The critical exponent $\nu$, on the other hand, appears to be largely independent of $k$ and is consistent with the critical exponent $\nu=4/3$ expected for a 1+1D NN model (\figref{fig:Percolation_Transition_Full}c, dotted black line).
These results suggest that all Haar-random \pwrk models with $k < \log_2 N$ fall into the same universality class as the local 1+1D model in the thermodynamic limit. Nevertheless, these results also demonstrate that the critical point $\pchaar$, a non-universal parameter, significantly increase with nonlocality $k$.

\section{Entanglement Transition in Clifford \pwrk circuits}
\label{sec:entanglement}

The nonlocal classical percolation network we studied in the previous section demonstrated that just a few additional layers of nonlocal interactions can significantly increase the critical point of the Haar-random \pwrk circuit. In the following sections we turn our attention to deterministic nonlocal circuits composed entirely of Clifford gates. We choose interaction gates $Q_{ij} = Q_{ij} = \cz_{ij} H_i H_j$ and show that the circuit's ability to withstand the destructive effects of local measurement is improved as a function of the nonlocality parameter $k$.

\begin{figure}
    \centering
    \includegraphics[width=1.0\columnwidth]{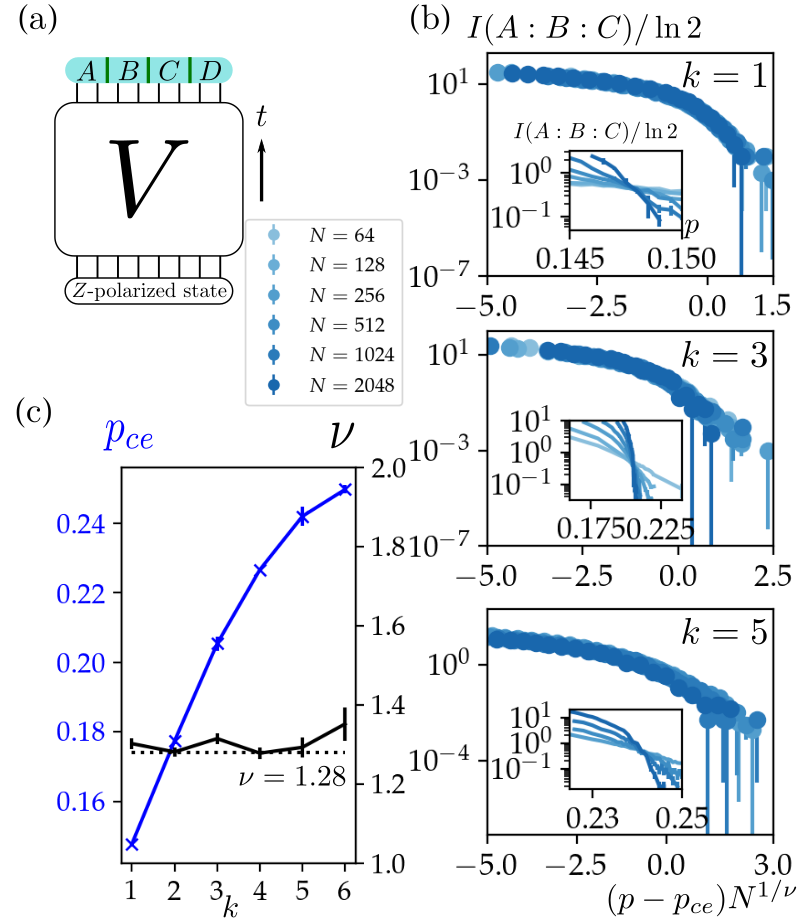}
    \caption{
        Measurement-induced entanglement transition in the \pwrk circuit family with Clifford gates.
        (a) The circuit is initialized with a separable pure input state, while the output at $T=8N$ is divided into four equal regions $A,B,C,D$. (b) Tripartite mutual information $I(A:B:C)$ for consecutive regions $A,B,C$ in the \pwrk circuit for 
        $k = 1,3,5$ (top to bottom) and system sizes $N = 64,\ldots,2048$ (light to dark blue).
        The critical point $p_{ce}$ is extracted from finite-size scaling for $N = 64,\ldots,2048$ (insets); only sufficiently large system sizes $N > 2^k$ are used to estimate the critical point. 
        Near $p_{ce}$ we observe a scaling collapse with the critical exponent $\nu\approx1.28$.
       (c) The critical measurement rate $p_{ce}$ (blue) increases significantly with $k$ 
       while the critical exponent $\nu$ (black) 
       is consistent with the 1+1D critical exponent (dotted black) to within statistical fluctuations for all values of $k$.
       Error bars are shown or are smaller than datapoints; lines are guides to the eye.
    }
    \label{fig:Entanglement_Transition}
\end{figure}

We begin our study of these nonlocal Clifford circuits by characterizing the entanglement at the output of the circuit as a function of the measurement rate $p$. We initialize the monitored \pwrk circuit with a pure, separable $z$-polarized state, and find a phase transition in the tripartite mutual information $I(A:B:C)$ between three equal-size consecutive regions $A,B,C$ of the output state as shown in Fig \ref{fig:Entanglement_Transition}a \cite{gullans2020dynamical}. As mentioned earlier, this quantity captures the connectivity of the circuit $V$ near the final time $t = T$.
The tripartite mutual information
\begin{align}
    I(A:B:C) = I(A,B) + I(A,C) - I(A,BC)
\end{align}
is defined in terms of the mutual information $I(A,B) = S^{(2)}_A + S^{(2)}_B - S^{(2)}_{AB}$, where $S^{(2)}_A = -\ln \tr{\rho_A^2}$ is the 
Renyi entropy of the subsystem $A$. Because the circuit consists entirely of Clifford gates, the Renyi entropy $S^{(2)}_A$ is always equal to the conventional von Neumann entropy $S_A = - \tr{ \rho_A \ln \rho_A}$ and we may therefore characterize the entanglement of the system entirely in terms of the Renyi entropies $S^{(2)}_A$.

To extract the entanglement critical point $p_{ce}$, we perform a finite-size scaling analysis similar to the previous section. In \figref{fig:Entanglement_Transition}b, we plot the tripartite mutual information as a function of measurement rate $p$ for various system sizes $N = 64, \ldots, 1024$ (light to dark blue). The critical point $p_{ce}$ is determined by the crossing point of the tripartite mutual information across system sizes as shown in the insets of \figref{fig:Entanglement_Transition}b. Only sufficiently large system sizes $N > 2^k$ are used to extract the critical point. Because the system is quasi-one-dimensional for finite $k < \log_2 N$ (see Appendix \ref{app:pwrkdimensionality}), near the critical point $p_{ce}$ we expect the tripartite mutual information to obey the universal scaling law
\begin{align}
    I(A:B:C) = f\left((p-p_{ce}) N^{1/\nu}\right)
\end{align}
where $f$ is a universal function and $\nu$ is the critical exponent of the correlation length $\xi$. After determining the critical point $p_{ce}$ we fit this scaling form to each curve in \figref{fig:Entanglement_Transition}b and use this to extract an estimate for the critical exponent $\nu$. The resulting estimates for $p_{ce},\nu$ allow us to collapse the original data down to a universal curve as shown in the main panels of \figref{fig:Entanglement_Transition}b \cite{gullans2020dynamical}.

We plot the resulting estimates for the critical point $p_{ce}$ and critical exponent $\nu$ in \figref{fig:Entanglement_Transition}c for $k = 1,\ldots,6$. Similar to our findings in the previous section, the non-universal critical point $p_{ce}$ increases significantly with $k$. The critical exponent, however, is nearly constant across $k$, and is consistent with the critical exponent  $\nu\approx1.28(2)$ found numerically for 1+1D NN models \cite{gullans2020dynamical} to within statistical fluctuations (\figref{fig:Entanglement_Transition}c).
This indicates that the measurement-induced transitions in these models likely fall into the same universality class as the completely local NN models.
Nevertheless, the fact that the critical point $p_{ce}$ increases with the nonlocality parameter $k$ demonstrates that just a few additional layers of nonlocal interactions can substantially improve a circuit's ability to retain complex many-body entanglement even in the presence of local measurements, similar to our observations in previous sections.

\section{Single-Qubit Purification in Clifford \pwrk Circuits}
\label{sec:purification}

\begin{figure}[h!]
    \centering
    \includegraphics[width=1.0\columnwidth]{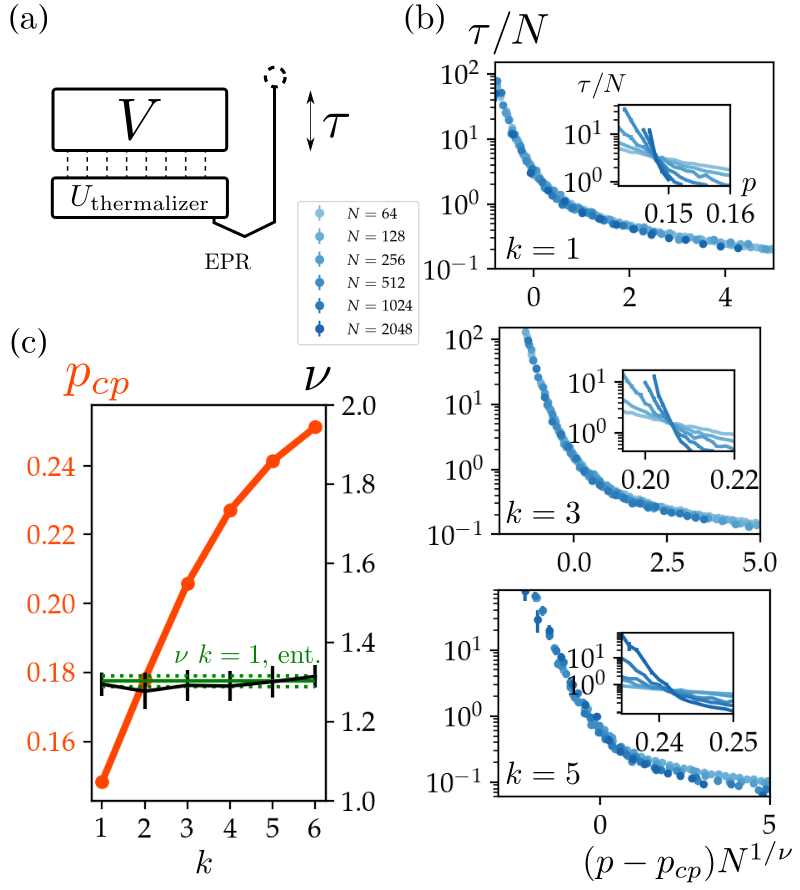}
    \caption{
        Single-qubit purification transition in the \pwrk circuit family with Clifford gates.
        (a) Schematic diagram for single-qubit purification.
        (b) Finite-size scaling for single-qubit purification time $\tau$ in the \pwrk circuit for $k=1,3,5$ (top to bottom) and system sizes $N = 64,\ldots,2048$ (light to dark blue).
        Main figures show scaling collapse with critical exponent $\nu \sim 1.30$ and dynamical critical exponent $z=1$ for $k=1,3,5$.
        The critical point $p_{cp}$ is determined by the crossing point of $\tau/L^{z=1}$ across system sizes $N > 2^k$ (insets).
        (c) The purification critical point $p_{cp}$ increases as a function of $k$, 
        closely mirroring the increase in the entanglement critical point $p_{ce}$ found in the previous section. 
        The critical exponent, on the other hand, agrees with the $\nu$ for $k=1$ found in the entanglement criticality
        in the previous section within the error (green solid line, with 1-sigma fluctuations indicated by the green dashed line).
        Error bars are shown or are smaller than the data points; lines are guides to the eye.
    }
    \label{fig:Purification_Transition_Points}
\end{figure}

Another measure of a circuit's robustness to measurements is the timescale $\tau$ required to purify a single qubit that has been entangled with the system \cite{gullans2020scalable,block2021measurement}. As mentioned earlier, this purification time captures the connectivity of the circuit $V$ between the initial and final times $t = 0,T$. Absent any measurements the circuit remains completely connected between $t = 0$ and $t = T$, and therefore the reference qubit remains entangled with the system forever. Conversely, carefully-placed projective measurements effectively tear the circuit apart and thereby destroy the entanglement between the system and qubit, causing the qubit to collapse into a pure state.
The typical timescale $\tau$ required for this purification process to occur can be used to characterize our circuit's ability to retain quantum information encoded in the initial state when subjected to noise. In particular, below a critical measurement rate $p < p_{cp}$ we expect the purification time to become extensive $\tau \sim N^z$, indicating that quantum information can be robustly stored in the circuit despite the presence of repeated projective measurements at a rate $p$. 

Although for nearest-neighbor circuits the purification critical point $p_{cp}$ coincides with the entanglement critical point $p_{ce}$, we emphasize that these transitions can generically be different, as has been pointed out in the literature \cite{gullans2020dynamical}.
We are especially interested whether highly nonlocal circuits support an intermediate phase in between the two critical points $p_{cp},p_{ce}$ where the reference qubit has been purified, 
but where there is nevertheless still volume-law entanglement in the final state. 

Here we study the possibility of such an intermediate phase in our nonlocal \pwrk circuits by determining the single-qubit purification time $\tau$.
To determine the purification critical point $p_{cp}$, we maximally entangle a single reference qubit $R$ with the system $Q$
and apply $t=4N$ layers of a unitary NN circuit to the system such that the qubit of information is scrambled deeply within the system \cite{block2021measurement} (\figref{fig:Purification_Transition_Points}a, $U_{\mathrm{thermalizer}}$).
Then the monitored \pwrk circuit is applied as illustrated in \figref{fig:Purification_Transition_Points}a. 
At the end of each timestep $t$ we compute the Renyi entropy $S_Q^{(2)}$ of the system. 
The single qubit purification time $\tau$ is the number of timesteps required for the Renyi entropy to vanish from its initial value of $S_Q^{(2)}(0) = \ln 2$. 

To extract the critical point $p_{cp}$ for each $k < \log_2 N$ we perform a finite-size scaling analysis as shown in \figref{fig:Purification_Transition_Points}b for $k = 1,3,5$. 
Similar to the analysis of previous sections, the critical point is determined by the location of the crossing point of $\tau(p)$ as the system size $N$ is varied, as shown in the insets of \figref{fig:Purification_Transition_Points}b.
The critical points $p_{cp}$ obtained from this analysis are plotted in \figref{fig:Purification_Transition_Points}c 
and grow significantly with $k=1,\dots,6$ in agreement with earlier analysis. In each of these cases the critical point $p_{cp}$ of the purification transition agrees with the entanglement critical point $p_{ce}$ within error bars.
This strongly suggests that for $k < \log_2 N$ the critical points $p_{cp}$, $p_{ce}$ are in fact identical and that there is no intermediate phase between the purification and entanglement phases \cite{gullans2020dynamical}.

As emphasized previously, the system is quasi-one-dimensional for finite $k < \log_2 N$, so we expect the purification time near the critical point to obey a universal 1+1D scaling law
\begin{align}
    \tau(p)= N^{z}f\left( (p-p_{cp})N^{1/\nu}\right)
    \label{eq:tauscaling}
\end{align}
where $z$ is the dynamical exponent and $\nu$ is the critical exponent of the correlation length $\xi$. Based on our findings in previous sections, we expect the purification transition studied here for $k < \log_2 N$ to be in the same universality class as the purification transition in 1+1D NN models, which are believed to be governed by a conformal field theory with dynamical exponent $z=1$ \cite{skinner2019measurement,gullans2020dynamical,block2021measurement}. We therefore assume $z = 1$ and use the scaling form \eqref{eq:tauscaling} to fit the data plotted in \figref{fig:Purification_Transition_Points}b. These fits yield estimates for the critical exponent $\nu$, which we plot in \figref{fig:Purification_Transition_Points}c. The resulting critical exponents $\nu=1.30 \pm 0.02$ are largely independent of $k$ and agree with the critical exponent of the entanglement transition of the 1+1D NN model. 
This consistent with previous results showing that the critical behaviour of the \pwrk with fixed $k < \log_2 N$ is in the same universality class as the conventional 1+1D MIPT transition.

\section{Quantum Error-Correcting Code Properties}
\label{sec:qecc}

Below the critical measurement rate $p < p_c$, the mixed phase is underpinned by a dynamically-generated quantum error-correcting code \cite{choi2020quantum,jian2020measurement,fan2020self,li2020statistical}. The improvement in the critical measurement rates $\pchaar, p_{cp}, p_{ce}$ as a function of $k$ observed in the previous sections suggests that circuits with highly nonlocal interactions generate improved quantum error-correcting codes in the mixed phase. An important characteristic of any QECC is its \emph{code distance}, which is the smallest number of single-qubit errors required to transform any code state into any other -- equivalently, the code distance is the minimal weight of all nontrivial logical operators.
Here we estimate the contiguous code distance $\dcode$ for our nonlocal Clifford circuits, and show that the improved robustness to local measurements observed in the previous sections also generates QECCs with improved code distance.

\begin{figure}
    \centering
    \includegraphics[width=1.0\columnwidth]{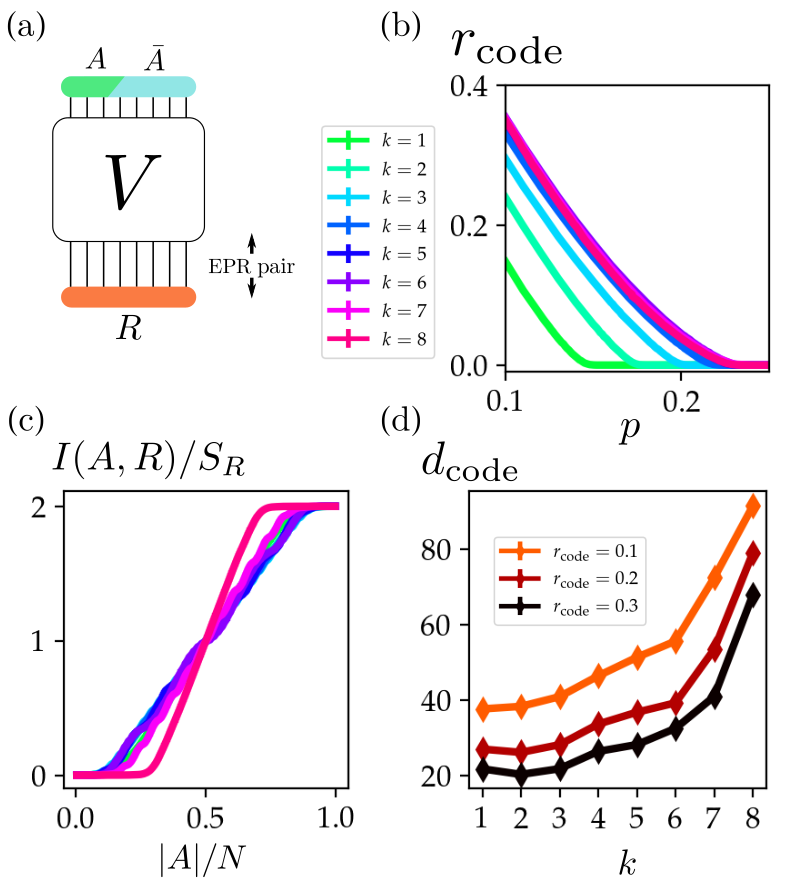}
    \caption{
        Code rate and code distance for the \pwrk Clifford circuit family at time $t = 8N$. 
        (a) To determine $\rcode,\dcode$, we examine the entropy $S_R^{(2)}$ of a maximally-entangled reference $R$ and the mutual information $I(A,R)$ between the reference and a subregion $A \subset Q$ of the output qubits. 
        (b) The code rate in the \pwrk circuit for $N = 256$ as a function of measurement rate $p$ improves significantly with $k$ (green through red).
        (c) Normalized mutual information as a function of subregion size for $p = 0.12$, deep in the mixed phase. 
        The effective contiguous code distance $\dcode$ is determined by finding the minimum size of the linear bipartition $|A|$ to have the 
        mutual information $I(A,\mathrm{Ref.})$ of $\ln 2$.
        (d) Code distance versus $k$ at fixed code rates $\rcode = 0.1,0.2,0.3$ (black, red, orange) for the system size $N=256$.
        Error bars are shown or are smaller than data points; lines are guides to the eye.
    }
    \label{fig:codeproperties}
\end{figure}

To characterize the QECC in the mixed phase, we maximally entangle the system $Q$ with a reference system $R$ and study the Renyi entropy $S^{(2)}_R$ of the reference as well as the mutual information $I(A,R)$ between the reference and a subset $A \subset Q$ of the output qubits as shown in \figref{fig:codeproperties}a. In the language of quantum error-correction, the Renyi entropy determines the \emph{code rate} $\rcode = S^{(2)}_R / N \ln 2$ of the \pwrk Clifford circuit, or the number of logical qubits that are encoded within the $N$-qubit system $Q$. We plot $\rcode$ as a function of $p$ in \figref{fig:codeproperties}b for $N = 256$ and find that the code rate substantially improves with nonlocality $k = 1,\ldots,8$ at any fixed measurement rate $p$, consistent with previous results.
From \figref{fig:codeproperties}b it is clear that for any fixed $k$, our choice of measurement rate $p < p_c$ below the critical point determines the code rate $\rcode$ of the underlying QECC.

We can also study the contiguous code distance $\dcode$ of the QECC in the mixed phase by analyzing the mutual information $I(A,R)$ between the reference $R$ and subregions $A \subset Q$ of the system as shown in \figref{fig:codeproperties}a \cite{li2020statistical}. For sufficiently small subregions $A$ in the mixed phase $p < p_c$, one generically finds vanishing mutual information $I(A,R) = 0$, indicating that the region $A$ contains no information about the reference $R$. In this circumstance we may safely discard any of the qubits in $A$ and still reliably recover the quantum information shared between the reference and the system. In the language of error correction, we can view the subregion $A$ as a set of qubits that has possibly been corrupted by errors. So long as $I(A,R) = 0$ we may simply discard these corrupted qubits and still recover the information contained in $R$. On the other hand, sufficiently large subregions $A$ will ultimately yield $I(A,R) > 0$, implying that sufficiently large errors can degrade the correlations between the system and reference. The crossover point $\magn{A}^*$ at which the mutual information becomes nonzero provides an estimate of the system's code distance $\dcode$ \cite{li2020statistical}. We review these arguments in more technical detail in Appendix \ref{app:qecc}.

To extract an estimate of the contiguous code distance $\dcode$, we plot the mutual information $I(A,R)$ as a function of the subregion size $\magn{A}/N$ and look for the crossover point $\magn{A}^*$ where the mutual information increase by one bit $\Delta I(A^*,R) = \ln 2$ (\figref{fig:codeproperties}c). The effective code distance is given by $\dcode=\langle \magn{A}^*\rangle$ where $\langle \dots \rangle$ is an average over different realizations of projective measurements in the circuit. We plot the resulting effective contiguous code distance as a function of $k$ in Fig. \ref{fig:codeproperties}d, which shows a striking improvement of $\dcode$ with increasingly nonlocal interactions $k$. Moreover, by tuning the measurement rate $p < p_c$ and the nonlocality parameter $k$ we can obtain quantum error-correcting codes with a variety of code rates and code distances. For fixed $k$ we find the expected tradeoff between code rate and code distance that is typical in quantum error-correcting codes. By increasing the nonlocality parameter $k$ we generate codes with significantly improved code distance $\dcode$ for any fixed code rate $\rcode$. In this sense, the nonlocal interactions for $k\sim \log_2 N$ substantially improve the quantum error-correcting code properties in the mixed phase.

\section{Fully Nonlocal and All-to-All Models}
\label{sec:nonlocalaa}

So far, we have considered \pwrk circuits at fixed $k < \log_2 N$ where the longest-range interactions $d_{\mathrm{max}} = 2^{k-1}$ have been strictly smaller than the system size $N$. In these cases we found that nonlocal interactions could substantially improve both the critical measurement rate $p_c$ and the code distance $\dcode$ in the mixed phase.
We now consider what happens in the `complete' \pwrk circuit with $k = \log_2 N$, where the longest-range interactions $d_{\mathrm{max}} = N/2$ are extensive, and show that the behavior radically changes relative to the cases previously studied. In particular, the complete \pwrk circuit without measurements is known to be a fast scrambler capable of generating system-wide volume-law entanglement after only $t_* \propto \log N$ interaction layers \cite{hashizume2021deterministic}. In this section we demonstrate that the fast scrambling dynamics in this circuit leads to markedly improved code properties, including a nearly extensive contiguous code distance $\dcode$. We also observe many similarities in this section between the complete \pwrk circuit and a maximally nonlocal random AA model, another known fast scrambler which has no spatial geometry whatsoever. These similarities highlight the central role played by fast scrambling in determining the physics of the mixed phase in these circuits, and suggest that fast scrambling circuits may be governed by the same universal physics near the measurement-induced critical point.

\begin{figure}
    \includegraphics[width=0.9\columnwidth]{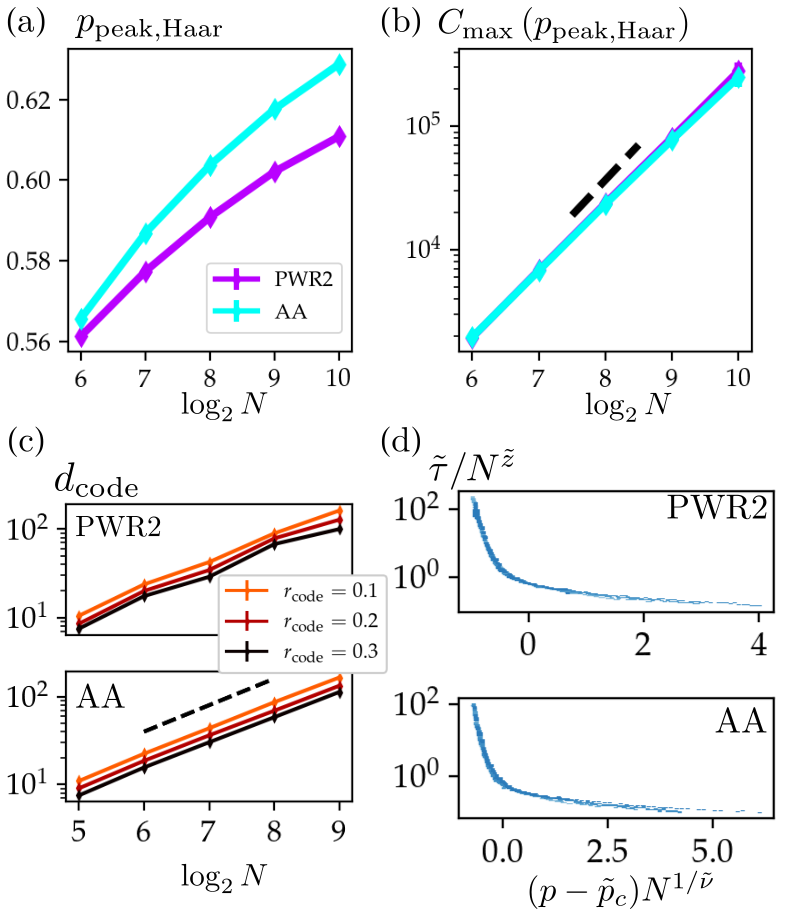}
    \caption{
        Maximally nonlocal `complete' \pwrk circuits with $k = \log_2 N$ and random AA circuits. 
        (a) Crossover points $p_{\mathrm{peak,Haar}}$ as a function of system size $N$ for the complete Haar-random \pwrk circuit 
        (purple) and for the random AA model (light blue).
        (b) The maximum cluster size $C_{\max}$ at the peak $p_{\mathrm{peak,Haar}}$ grows as a power law $N^{\mu}$ with system size for the complete \pwrk circuit (purple) and the AA circuit (light blue), with exponents 
        $d_f = 1.809\pm0.005$ and $d_f = 1.786\pm0.005$ respectively.
        Black dashed line shows the fractal dimension, $d_f=91/48$, of the cluster at the critical point of two-dimensional system.
        (c) The code distance for the complete \pwrk Clifford circuit (top) and the random AA Clifford circuit (bottom) for code rates $\rcode=0.1,0.2,0.3$ (black, red, orange). In both cases, the code distance is nearly extensive, scaling like $\dcode \propto N^{\beta}$ with $\beta=0.96\pm0.04$ (\pwr) and $\beta=0.97\pm0.02$ (AA).
        (d) Scaling collapse of the normalized single-qubit purification time $\tilde{\tau}(p)$ in the complete \pwrk Clifford circuit (top) and the random AA Clifford circuit (bottom). In both cases the data exhibits strong collapse according to the nonstandard scaling form in Eq. \eqref{eq:logscalingcollapse}.
    }
    \label{fig:completealltoall}
\end{figure}

We first study the complete \pwrk circuit with Haar-random gates, and numerically simulate the corresponding classical percolation network similar to section \ref{sec:percolation}. In that prior analysis we computed the Binder cumulant $b(p)$ of the network and estimated the critical point by fixing the parameter $k$ and performing a finite-size scaling analysis in the system size $N$. In the present case, finite-size scaling is difficult to define consistently because the connectivity $k = \log_2 N$ of the graph is coupled to the system size $N$. Instead, we estimate a finite-size crossover point by computing the susceptibility
\begin{equation}
    \chi(p) = \langle C^2_{\max}(p)\rangle - \langle C_{\max}(p)\rangle^2
\end{equation}
of the resulting classical percolation network. In the thermodynamic limit $N \rightarrow \infty$ the susceptibility diverges at the critical point; here we estimate the finite-size transition point by locating the peak $p_{\mathrm{peak,Haar}}$ of the susceptibility $\chi(p)$ for each $N$.

The resulting finite-size crossover points are plotted in \figref{fig:completealltoall}a, where we compare to the crossover points for a random AA model analyzed using the same methods. The crossover points noticeably increase with $N$ in both cases, and should therefore be considered as finite-size crossover points and not bona fide critical points. We also plot the value of the maximum cluster size at the peak $C_{\max}(p_{\mathrm{peak,Haar}})$ in \figref{fig:completealltoall}b, and find that it increases as a power law with system size 
$C_{\max}(p_{\mathrm{peak,Haar}}) \propto N^{\mu}$ with fractal dimension $d_f = 1.809\pm0.005$ for the complete \pwrk circuit and $1.786\pm0.005$ for the random AA circuit. The similarity in the fractal dimension $d_f$ between these two circuits suggests that they may be governed by the same universal physics near the critical point. Moreover, the fractal dimension in both cases differs from the analytical prediction $d_f = 91/48 \approx 1.896$ obtained from the two-dimensional percolation universality class. This suggests that there is an abrupt change in the universality class from the $k<\log_2 N$ circuits to the complete \pwrk circuit.

Next, we turn our attention to circuits composed of Clifford circuits with two-qubit gates $Q_{ij} = \mathrm{CZ}_{ij} H_i H_j$ and study the quantum error-correcting codes that support the mixed phase. Again, we find striking similarities with the random AA circuit, suggesting that these models may be governed by the same universal physics.  Using the methods discussed in Sec. \ref{sec:qecc}, we extract the code distance $\dcode$ of the QECC in the mixed phase at fixed code rate $\rcode$. We plot the results in \figref{fig:completealltoall}c for code rates $\rcode = 0.1,0.2,0.3$. Linear fits indicate a nearly extensive code distance, where $\dcode \propto N^{\beta}$ with $\beta=0.98\pm0.05$ for the \pwr circuit and $0.96\pm0.01$ for the random AA circuit. 

This demonstrates that the fast scrambling dynamics generated by the maximally nonlocal \pwrk model generates quantum error-correcting codes with excellent properties deep in the mixed phase. The nearly extensive code distance $\dcode \sim N$ found here is in agreement with the extensive code distance expected at very long times $T \gg \exp{(N \rcode)}$ in random 1+1D circuits \cite{li2020statistical}, as well as with results from short random circuits which achieve codes at the Gilbert-Varshamov bound \cite{brown2013short,brown2015decoupling}. In this sense, our \pwrk circuits achieve the same QECC properties that are known to be accessible in random circuits. A key distinction is the fact that the sparse nonlocal circuits studied here are deterministic and can be implemented in near-term experiments with Rydberg arrays \cite{hashizume2021deterministic} or in cavity QED experiments \cite{bentsen2019treelike,periwal2021programmable}. One may be concerned that the inherent randomness in the projective measurements inside the non-unitary circuit $V$ would preclude the practical use of these codes in realistic experiments because the encoding circuit changes on every experimental run depending on the measurement outcomes. However, we demonstrate in Appendix \ref{app:subpracticalerrorcorrection} how this problem can be overcome for Clifford \pwrk circuits by leveraging a parallel classical Clifford simulation alongside the quantum experiment. This procedure allows these codes to be used in practical applications.

Finally, we study the single-qubit purification time $\tau$ in the complete \pwrk circuit with $k = \log_2 N$. 
Instead of the conventional scaling law \eqref{eq:tauscaling} near the critical point, here we empirically find strong scaling collapse of $\tau$ only after normalizing by the number of interaction layers $k = \log_2 N$ within each even (or odd) block.
This leads to a nonstandard scaling law
\begin{equation}
    \tau / \log_2 N=\tilde{\tau}(p) = N ^{\tilde{z}} \tilde{f}\left( (p-\tilde{p}_c)N^{1/\tilde{\nu}} \right),
    \label{eq:logscalingcollapse}
\end{equation}
which yields strong data collapse for both the complete \pwr model and the random AA model as shown in \figref{fig:completealltoall}d.
This suggests that the critical point in the fast-scrambling limit may be governed by a logarithmic scaling law \eqref{eq:logscalingcollapse} as opposed to the conventional scaling \eqref{eq:tauscaling} that governs the phase transition in general short- and long-range models \cite{block2021measurement,minato2021fate}.
The origin of this scaling behaviour is an interesting topic for further investigation. 

\section{Discussion and Outlook}
\label{sec:conclusion}

In this paper we studied the role of sparse nonlocal interactions in protecting quantum information against the destructive effects of local measurements. We characterized the protection afforded by these nonlocal interactions by numerically studying measurement-induced phase transitions in a class of sparse, nonlocal \pwrk circuits with tunable nonlocality $k$. These studies demonstrated that non-universal properties of the transition, such as the critical points $\pchaar,p_{ce},p_{cp}$, can be significantly improved by the nonlocality of the interaction graph. At the same time, we found that the universal properties of the transition such as the critical exponent $\nu$ are largely unaffected by the presence of these nonlocal interactions so long as $k < \log_2 N$. In the language of the renormalization group, it appears that sparse nonlocal interactions are an irrelevant perturbation to the 1+1D fixed point. It would be interesting to make this notion of irrelevancy more precise within a renormalization group framework.

While the phase transition for fixed $k < \log_2 N$ appears to be governed by the same universality class as the 1+1D model, we find that the properties of the system deep in the mixed phase can change considerably in the presence of nonlocal interactions. In particular, we have demonstrated that the code distance $\dcode$ (at fixed code rate $\rcode$) can be significantly improved by the presence of sparse, nonlocal interactions. By varying the parameter $k$, we can thus tune between the code properties of a NN circuit and the improved code properties of an AA circuit.

We also studied `complete' \pwrk circuits with $k = \log_2 N$ and observed that these circuits are closely comparable to known fast scramblers such as the random AA circuit. Although a proper finite-size scaling analysis is not available for these circuits, we numerically studied finite-size crossover points and showed that the critical properties of the complete \pwrk circuit are markedly similar to those of a random AA circuit. We also studied the code properties of the complete \pwrk and AA Clifford circuits, and found codes with nearly extensive code distance $\dcode \propto N^{\beta}$ with $\beta \approx 1$ in both cases. Finally, we observed a non-standard scaling collapse in the single-qubit purification time $\tau$ for both models, another signal that the critical properties of these models may be described by the same universality class. In future work we hope to make these connections more explicit.

One of the most exciting prospects for engineering fast scrambling circuits in the lab is the ability to generate good quantum error-correcting codes using the fewest gates possible. Here we have demonstrated that deterministic fast scrambling circuits with local projective measurements are capable of generating good quantum error-correcting codes, and that the properties of these codes can be easily tuned. It is natural to ask whether one could reasonably use these monitored circuits to generate useful codes in near-term experiments. One outstanding problem from an experimental perspective, however, is the post-selection problem. Because quantum projective measurements are inherently probabilistic, one can only force a particular outcome by repeating the protocol ad infinitum until the correct measurement record is obtained. This necessarily requires the experiment typically be repeated an exponentially large number of times, which is impractical. In the case of Clifford circuits it is known in principle how to apply feedback to effectively correct for undesired measurement outcomes \cite{noel2021observation}. It would be interesting to apply these ideas to our fast scrambling Clifford circuits to understand whether single-shot experiments with suitable feedback can be used to generated useful quantum error-correcting codes in near-term experiments with cold neutral atoms or trapped ions. The data for this manuscript is available in open access at \cite{hashizume2021data}.

\begin{acknowledgments}
We thank Hans Peter B\"uchler and Nicolai Lang for helpful discussions. GSB is supported by the DOE GeoFlow program (DE-SC0019380). Work at the University of Strathclyde was supported by the EPSRC Programme Grant DesOEQ (EP/P009565/1), the EPSRC Quantum Technologies Hub for Quantum Computing and simulation (EP/T001062/1), the European Union’s Horizon 2020 research and innovation program under grant agreement No. 817482 PASQuanS, and AFOSR grant number FA9550-18-1-0064.
\end{acknowledgments}

\bibliography{References}

\appendix

\section{Haar-Random Circuits and Bond Percolation}
\label{app:haarpercolation}

\subsection{Percolation Mapping}
\label{app:subhaarpercolation}

\begin{figure}
  \includegraphics[width=\columnwidth]{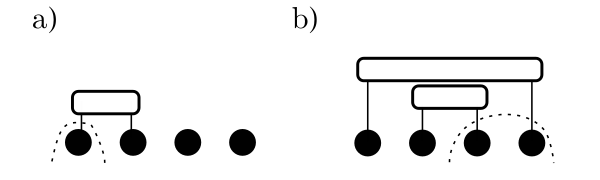}
    \caption{
        (a) Haar-random two-site gates typically generate nearly maximal entanglement. In this case the zeroth Renyi entropy $S_0$ of site $i = 0$ is proportional to the minimum number of cuts to required to isolate it, $S_0 / \ln 2= 1$.
    (b) Haar-random gates applied between sites $i,j=1,2$ and sites $i,j=0,3$.
    In this case $S_0$ takes a value of $S_0 / \ln 2 = 2$ for the subsystem consisting of sites $i = 0,1$.\label{FigAppHaar}}
\end{figure}

The zeroth order Renyi entropy is $S_0=\log M$,
where $M$ is the rank of the reduced density matrix $\rho_A$ of the subsystem $A$.
When an independent qubit is entangled with $A$, the rank of $\rho_A$ increases by a factor of $2$. 
A projective measurement, on the other hand, isolates the measured qubit from the rest of the system, typically undoing the effect of the entangling gates applied previously to the measurement.
In the tensor network picture, the calculation of $S_0/\ln 2$ maps to a classical minimal cut problem.
In this picture the zeroth Renyi entropy is given by the minimum number of legs one must cut in order to isolate the qubits in $A$ (see Fig. \ref{FigAppHaar}).

By considering the two extreme limits of the probability of the projective measurement $p$, 
one finds that in the limit of $p=0$ the initial state and final state is guaranteed to be connected, and 
in the limit of $p=1$, there is no connection between them.
Between these two limits there must be a critical probability $p=\pchaar$ which separates the two phases.
For $p<\pchaar$, there always exist a path between qubits in the initial state and the final state,
and for $\pchaar<p$, they are independent.

If two qubit gates are taken from the Haar-random distribution and applied to a pair of independent qubits,
then two qubit gates are almost always guaranteed to be entangled.
In this case, the circuit can be drawn as a tensor network diagram consists of entangling tensors (Haar-random gates) 
which act on the pairs of qubits accordingly to the interaction graph, with the legs being cut with probably $p$ 
due to the projective measurements.
This is equivalent to the bond-percolation with the bond occupation probability $q=1-p$.
The critical point, $\pchaar$, therefore, can be determined by solving the classical bond percolation problem 
of the mapped network. 

\subsection{Numerical Method for Simulating Bond Percolation}
\label{app:subnewmanziff}

The Newman-Ziff algorithm \cite{newman2001fast} is used for simulating the critical behaviour of 
bond percolation.
This algorithm calculates the various observables at different values of the bond occupation probability $q=1-p$ 
in a computational time which scales only linearly with the number of bonds $M$ in the network.

The algorithm proceeds by generating random configurations of a given network with $m=1,2,\dots,M$ bonds being occupied;
where this is done by adding one random bond at each step starting from an empty network.
At each step of this process, one can keep track of the sizes of the connected components (clusters) in the network 
with the Union-Find algorithm \cite{hopcroft1973set,tarjan1975efficiency}.
Observables such as cluster sizes and their moments can then be stored at the end of each step of the algorithm.
The observable $Q$ as a function of the bond occupation probability $q$
can then be calculated by taking the microcanonical ensemble of the different configurations: 
\begin{align}
    Q(p) 
    = \sum_{m} 
    B(m,M,q) Q_m
    = \sum_{m} 
    \begin{pmatrix}
        M \\
        m
    \end{pmatrix}
    q^{m} q^{M-m} Q_m
\end{align}
where $Q_m$ is the value of observable $Q$ with $m$ occupied bonds and 
$\begin{pmatrix}M \\ m\end{pmatrix} =\frac{M!}{m!(M-m)!}$ is the usual binomial coefficient.

The calculation of the binomial distribution $B(m,N,p)$ is numerically unstable because 
one requires evaluations of the factorials of large numbers and addition of numbers which differ largely 
in the order of the magnitudes.
Here we adopt a more stable method for evaluating the microcanonical ensemble introduced by Newman and Ziff \cite{newman2001fast}.
If we normalize the binomial distribution by its peak value $m_{\max}=pM$, this normalized binomial distribution $\tilde{B}(m,M,p)$ is defined recursively as
\begin{align}
    \tilde{B}(m,M,q) = 
    \begin{cases}
        \tilde{B}(m-1,M,q)\frac{M-m+1}{m}\frac{q}{1-q} &(m > m_{\max}) \\
        \tilde{B}(m+1,M,q)\frac{m+1}{M-m}\frac{1-q}{q} &(m < m_{\max})
    \end{cases}
\end{align}
Therefore the observable of interest can be calculating the following
\begin{align}
    Q(q) = \frac{\sum_{m'}\tilde{B}\left( m',M,q \right) Q_m}{\sum_{m'}\tilde{B}\left( m',M,q \right) }.
\end{align}

\subsection{Renormalization Group Solution for full PWR2}
\label{app:Renormalization}

The PWR2 circuit with size $N_0$ can be constructed from two sub-systems of size $N_0/2$.
This can be done by interleaving the degrees of freedom of the two sub-systems and 
coupling them together with the nearest neighbor interactions. 

\begin{figure}[t!]
  \includegraphics[width=\columnwidth]{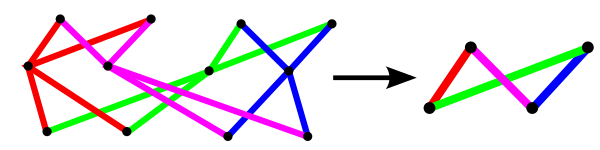}
  \caption{
    Block-decimation procedure for the $k=2$ \pwrk circuit.
    The entire structure on the left is reduced down to a single ribbon-like structure on the right via the renormalization
    transformation. 
  }
  \label{fig:Renormalization}
\end{figure}

This self-similar structure allows us to define a renormalization transformation.
This transformation consists of two steps:
First, a large block as depicted in \figref{fig:Renormalization}~(left),
is renormalized to a ribbon-like structure in \figref{fig:Renormalization}~(left).
Here the renormalized is done by collapsing the bonds with different colors in \figref{fig:Renormalization}~(left) 
to the bonds of the corresponding colors in \figref{fig:Renormalization}~(right).
This step reduces the size of the vertical dimension by half.
In order to reduce the size of the horizontal direction, two of the ribbon-like structure is merged into one.
After these steps, the size of the network of $N/2$ by $T$ is reduced to $N/4$ by $T/2$,
where $N$ is the number of qubits and $T$ is the number of layers in the original circuit.
It is important to note that from \figref{fig:Renormalization}~(left), 
non-vanishing amount of loops are always present for $4<N$ between each layers. 

This implies that the model is fundamentally different to hypercubic/tree-like models discussed in \cite{nahum2020measurement}. 

Let $q=1-p$ be the probability of a bond being present in the network.
The renormalized probability $R(q)$ in terms of the probability which a ribbon structure
(\figref{fig:Renormalization}, right) is spanned ($Q_{\mathrm{ribbon}}$) is 
\begin{align}
    R(q) &= Q_{\mathrm{ribbon}}^2 + 2Q_{\mathrm{ribbon}}(1-Q_{\mathrm{ribbon}}).
\end{align}
Now we define the condition of the ribbon to be spanned when there exists at least one bond coming out of each the nodes.
In terms of the probability of at least a bond being spanned $q_{\mathrm{bond}}$, $Q_{\mathrm{ribbon}}$ is 
\begin{align}
    Q_{\mathrm{ribbon}} &=q_{\mathrm{bond}}^{4}
  +4q_{\mathrm{bond}}^{3}(1-q_{\mathrm{bond}}) 
  +2q_{\mathrm{bond}}^{2}(1-q_{\mathrm{bond}})^2.
\end{align}
Finally, the probability of the bond being spanned, $q_{\mathrm{bond}}$, as a function of $q$ is 
\begin{align}
    q'&=q^4+4q^3(1-q)+4q^2(1-q)^2.
\end{align}
The equation $R(q^*)=q^* $ has a non-trivial fixed point solution in the real domain of $0<q^*<1$, 
which is $q^*\approx0.335\dots$ or equivalently $p^*\approx0.665\dots$.


\section{Dimensionality of \pwrk Model with Fixed $k$}
\label{app:pwrkdimensionality}

\begin{figure}
   \includegraphics[width=0.4\textwidth]{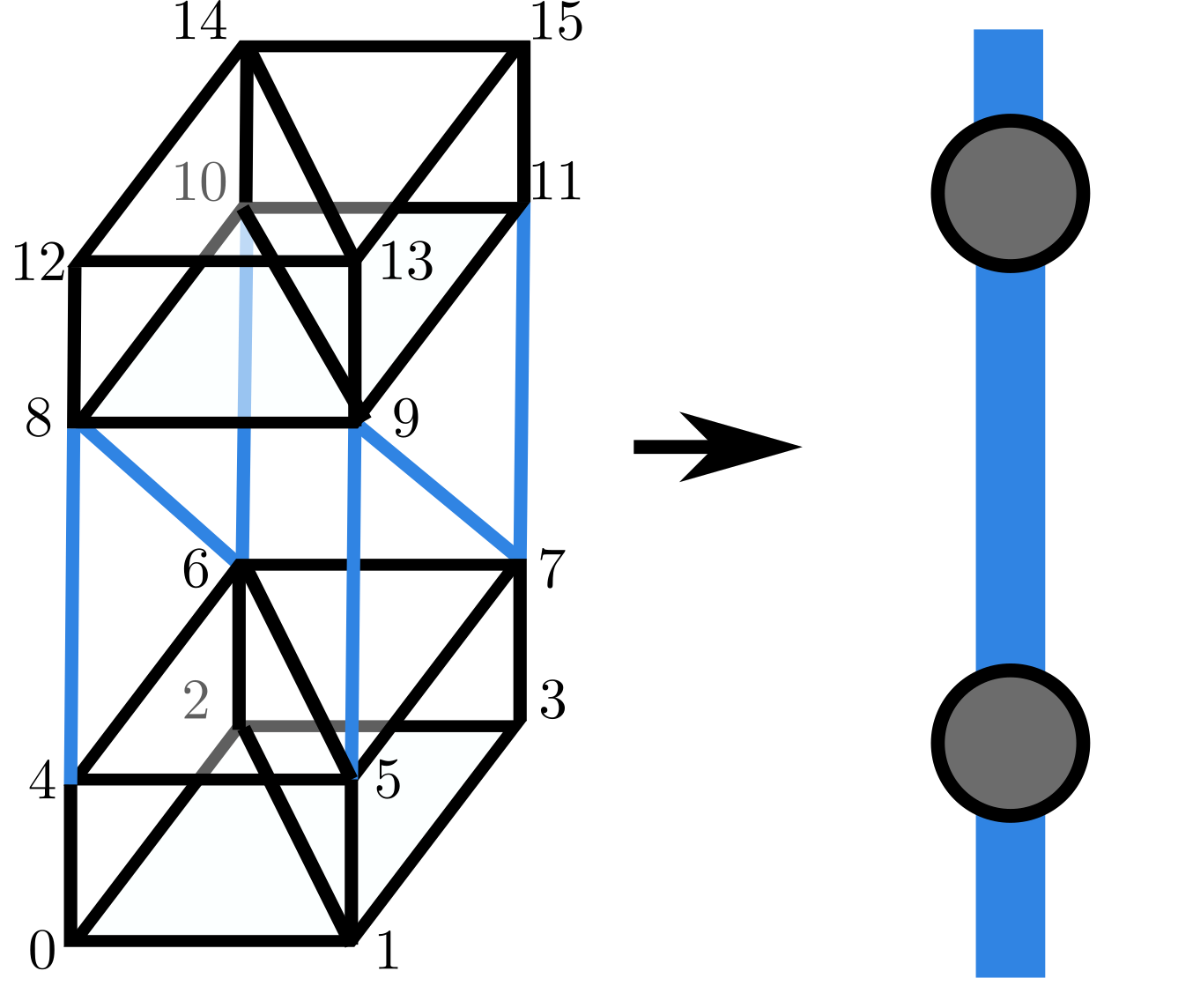}
   \caption{The decomposition of the interaction graph of \pwrk with $k=3$ into the chain of hypercube-like interactions.
   Two sets of $2^{k}$ contiguous sites are interacting through \pwrk interaction forming a hypercube-like geometry 
   (black lines, left).
   The sites on the Two neighboring $k-1$ dimensional faces of the hypercubes are coupled through interactions depicted in blue.
   This can be interpreted as chain of sites with each having $D=k$ dimensional hypercube-like degrees of freedom (right).
   \label{fig:ChainofHypercubes}}
\end{figure}
The interactions within a contiguous region of size $N_{\sub}=2^{k}$ in \pwrk circuit with fixed $k$ 
can be embedded into a $k$-dimensional cubic lattice.
The rest of the bonds connect two neighboring $k-1$ dimensional faces of the hypercube. 
The resulting geometry is a quasi-one-dimensional chain of $N/2^{k-1}$ sites where each site hosts a $k$-dimensional 
hypercubic degree of freedom
as shown in \figref{fig:ChainofHypercubes}.
In sections III-IV, we observed both critical exponent $\nu$ and dynamic exponent $z$ that are consistent 
with the nearest neighbor models for fixed $k$ \pwrk model.
This chain-like critical behavior comes from the global chain-like structure which the models have for sufficiently large enough $N$ 
(\figref{fig:ChainofHypercubes} left). 

\section{Clifford Circuits}
\label{app:cliffords}

\subsection{Stabilizer States}
\label{app:substabilizer}

The stabilizer formalism is a powerful tool for understanding a special class of many-body quantum states and quantum error-correcting codes whose dynamics can be efficiently simulated on a classical computer \cite{gottesman1998heisenberg,aaronson2004improved}. Consider the Pauli group $\mathcal{P}(Q)$ of all Pauli strings acting on a set of qubits $Q$. We define an Abelian \emph{stabilizer} subgroup $\mathcal{S} \leq \mathcal{P}(Q)$ generated by a set $\mathcal{G} = \{g_1,g_2,\ldots,g_m\}$ of linearly-independent, mutually-commuting Pauli strings $[g_{\ell},g_{\ell'}] = 0 \ \forall \ell,\ell'$. A stabilizer group with $m \leq N = \magn{Q}$ independent generators has order $\magn{\mathcal{S}} = 2^m$. Given a stabilizer group $\mathcal{S}$ we define a \emph{stabilizer state} (also called a \emph{code state})
\begin{equation}
    \rho_Q(\stabs) = 2^{-\magn{Q}} \sum_{g \in \mathcal{S}} g
    \label{eq:codestate}
\end{equation}
which is the unique density matrix stabilized by all elements $g \in \mathcal{S}$. If we specify a `complete' set of $m = N$ stabilizers, then \eqref{eq:codestate} is a rank-1 projector onto the unique pure state $\ket{\Sigma}$ stabilized by the group $\mathcal{S}$ (i.e. $g\ket{\Sigma} = +\ket{\Sigma}$ for all $g \in \mathcal{S}$).

Quantum error correction is particularly easy to understand in the stabilizer formalism. Suppose we prepare a code state \eqref{eq:codestate} defined by a stabilizer group $\mathcal{S}$, and consider disturbing it with an error operator $e \in \mathcal{P}(Q)$. There are three possibilities: first, if $e \in \mathcal{S}$ then the error is \emph{trivial} because it leaves the state \eqref{eq:codestate} unchanged. Second, if $e \notin \mathcal{S}$ fails to commute with one or more stabilizers $g$ then this is a \emph{detectable} error because we can detect (and subsequently correct) the error by measuring the stabilizer $g$. The final possibility is if $e \notin \mathcal{S}$ but $e \in \mathcal{C}(\mathcal{S})$ is in the centralizer of $\mathcal{S}$. These errors are \emph{undetectable} because they modify the quantum state but cannot be detected by measuring any of the stabilizers $g$. The collection of all undetectable errors gives the logical operator group $e \in \mathcal{L} \equiv \mathcal{C}(\mathcal{S})/\mathcal{S}$.

\subsection{Classical Simulation of Clifford Circuits}
\label{app:subclifford}

Clifford group on $N$ qubits, $\mathcal{C}^{N}$, is a group formed by unitary operators which transforms the elements of 
$N$ qubit Pauli group to other elements in the same group.
The action of those unitaries on a stabilizer of $\mathcal{O}_{\ket{\Sigma}}$ 
therefore transforms the stabilizer to other stabilizer $\mathcal{O}_{\ket{\Sigma'}}$.
This transformation is equivalent to the transformation of state $\ket{\Sigma}$ to $\ket{\Sigma'}$.
Therefore, the evolution of a stabilizer state under the unitaries from the Clifford group can be 
tracked by keeping track of how the initial stabilizer transforms.

The evolution with unitaries from the Clifford group 
is proven to be able to be computed classically in polynomial time 
by Gottesman and Knill \cite{gottesman1998heisenberg,aaronson2004improved}.
This is done with a binary matrix $M$ of size $N$ by $2N$. 
In this representation, Pauli strings in the stabilizer are mapped to each rows of $M$
and the actions of Clifford gates are mapped as series of logical operations \cite{aaronson2004improved}.

In this simulation a Pauli strings are mapped to a row in the following fashion.
A Pauli operators which corresponds to the $l$\textsuperscript{th} qubit is encoded by mapping the number 
of Pauli-$X$ to $l$\textsuperscript{th} column and the number of Pauli-$Z$ to $N+l$\textsuperscript{th} column.
If $l$\textsuperscript{th} operator is Pauli $Y$, then because of the identity, $Y=iXZ$, 
the values of the $l$\textsuperscript{th} and $N+l$\textsuperscript{th} columns is registered as $1,1$.
If $l$\textsuperscript{th} operator is an identity, both of the values are registered as $0$.

To keep track of the evolution by the unitaries in the Clifford group, 
it is useful to identify the transformation rule of the binary matrix for the generators of the Clifford group.
The generator of $N$-qubit Clifford group are Hadamard ($H$), Phase ($P$) 
and Controlled-NOT (C-NOT) gates \cite{gottesman1998heisenberg,aaronson2004improved,selinger2015generators}.
Hadamard gate acting on a site, $l$, maps operators $Z_l$ to $X_l$ and $X_l$ to $Z_l$ while keeping $Y_l$ to be $Y_l$.
This is equivalent to swapping the columns $l$ and $N+l$.
Phase gate acting on a site, $l$, on the other hand, maps $X_l$ to $Y_l$ and $Y_l$ to $X_l$ while keeping $Z_l$ unchanged. 
This is equivalent to setting the column $N+l$ as the modulo $2$ sum of columns $l$ and $N+l$.
Let qubit $l$ be a control and qubit $m$ be the target, then the Controlled-NOT gate acting on those qubits
flips the target basis whenever the state of the $l$ is $\ket{1}$. 
By transforming all the elements in the two-qubit Pauli group,
one finds that there are only four non-trivial rules, which are :
$X_lI_m$ to$X_lX_m$, $I_lX_m$ to $I_lX_m$, $Z_lI_m$ to $Z_lI_m$, and $I_lZ_m$ to $Z_lZ_m$.
Since $H$, $P$, and C-NOT gates are the generators of the Clifford group, one may systematically generate all the operators in 
$n$-qubit group by following the algorithms such as the one proposed by Selinger \cite{selinger2015generators}.

\subsection{Entanglement Entropy}
\label{app:subentropy}

The stabilizer is equivalently the density matrix of the stabilizer state because the stabilizer is a projector 
which projects the state onto the stabilizer state.
With the density matrix $\rho_Q(\mathcal{S})$, entanglement entropy of a subregion $A$ 
can be calculated \cite{nahum2017quantum,gullans2020dynamical}.

The reduced density matrix on the subregion $A$, $\rho_A$ ($A\in Q$) is obtained by tracing out $\bar{A}$.
This is equivalent to tracing out the Pauli operators which belong to $\bar{A}$.
However, Pauli operators are traceless, therefore 
\begin{align}
    \rho_{A} = \mathrm{Tr}_{\bar{A}} \{ \rho_Q \} 
    = \frac{2^{|\bar{A}|}}{2^{Q}}\sum_{g_A\in\mathcal{G}_A} g_{A} = \frac{1}{2^{|A|}}\sum_{g_A\in\mathcal{G}_A} g_{A},
\end{align}
where $\mathcal{G}_A\in\mathcal{G}$ is a set of all $g$ where trace over $\bar{A}$ is non-zero.
\cite{nahum2017quantum}.
Let $N_A$ be the number of linearly independent Pauli strings that generate $g_A$, then 
$\sum_{g_A\in\mathcal{G}_A} g_A$ is also a projector of rank $2^{|A|-N_A}$ because this 
projects out the $-1$ eigenstates of its generators. 
The von Neumann or Renyi entropy $S_A$ is 
\begin{align}
    S_A = \left(|A| - N_A \right)\ln 2
\end{align}
From the identities $2^{N_A+N_{\bar{A}}} = 2^{N}$ and $N_{\bar{A}}=\mathrm{rank}_{\mathrm{binary}} M_{\bar{A}}$,
where $M$ is the binary matrix of the stabilizer state, $\mathrm{rank}_{\mathrm{binary}}$ is the binary rank,
and $M_{\bar{A}}$ is the matrix which its columns corresponds to the Pauli operators of region $\bar{A}$ in $M$,
\begin{align}
    S_A = \left(\mathrm{rank}_{\mathrm{binary}}(M_{\bar{A}}) - |\bar{A}| \right)\ln 2.
\end{align}
Using $S_A=S_{\bar{A}}$, we obtain
\begin{align}
    S_A = \left(\mathrm{rank}_{\mathrm{binary}}(M_A) - |A|\right)\ln 2.
\end{align}

\section{Finite-Size Scaling}
\label{app:finitesizescaling}

\subsection{Finite Size Scaling of the Percolation Critical Point for $k=1,\dots,6$}
\label{app:scalingcollapseHaar}

\begin{figure*}[t!]
  \includegraphics[width=0.95\textwidth]{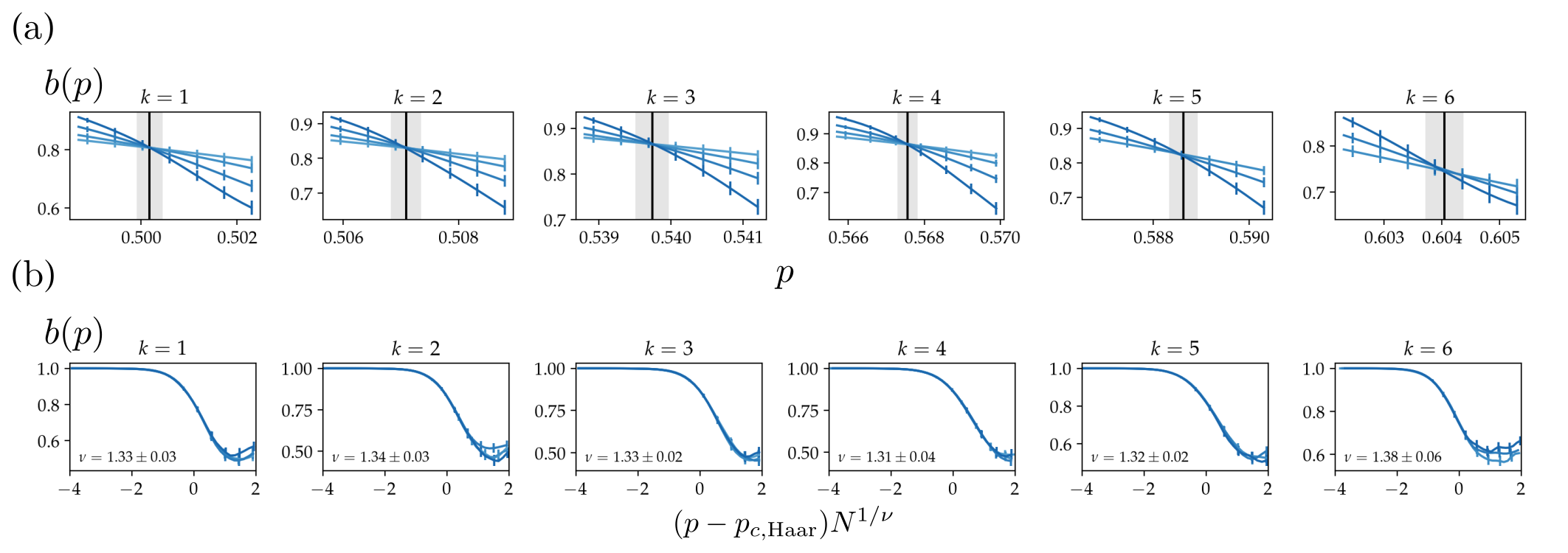}
  \caption{
      (a) The binder cumulant $b(p)$ of the percolation network of \pwrk for $k=1,2,\dots, 6$
      for the system sizes $N=2^{8},2^{9},2^{10},2^{11}$ (light to dark blue).
      The solid vertical lines are the estimated values of the $\pchaar$ and the shaded regions are the corresponding $1$-sigma errors.
      (b) Collapsed Binder cumulant for $k=1,2,\dots6$
      for the system sizes $N=2^{8},2^{9},2^{10},2^{11}$ (light to dark blue).
      \label{App:FSPercolation}
  }
\end{figure*}

Finite-size scaling analysis for the percolation transition of the \pwrk circuit for $k=1,3,5$
was presented in Fig. 2~b of the main text.
Here we show the finite-size scaling and the crossings of the binder cumulants for all values of $k=1,2,\dots,6$
for the system sizes $N=2^{8},2^{9},2^{10},2^{11}$.

The error bars are estimated by taking the standard error of the 4000 realizations of the Newman-Ziff algorithm (see Appendix \ref{app:haarpercolation}).
The critical points and their errors are estimated by computing the average intersection points 
of the curves with 5000 different realizations of the fluctuations added to each of the points.
Fluctuations are drawn from the appropriate distribution with the standard error as the standard deviation at each point
(\figref{App:FSPercolation}~a). 
The critical exponents and their errors are also estimated by collapsing the 5000 different realizations of the fluctuations
added from the same distribution (\figref{App:FSPercolation}~b).

\subsection{Finite Size Scaling of the Entanglement Critical Point for $k=1,\dots,6$}
\label{app:scalingcollapseEnt}

\begin{figure*}[t!]
  \includegraphics[width=0.95\textwidth]{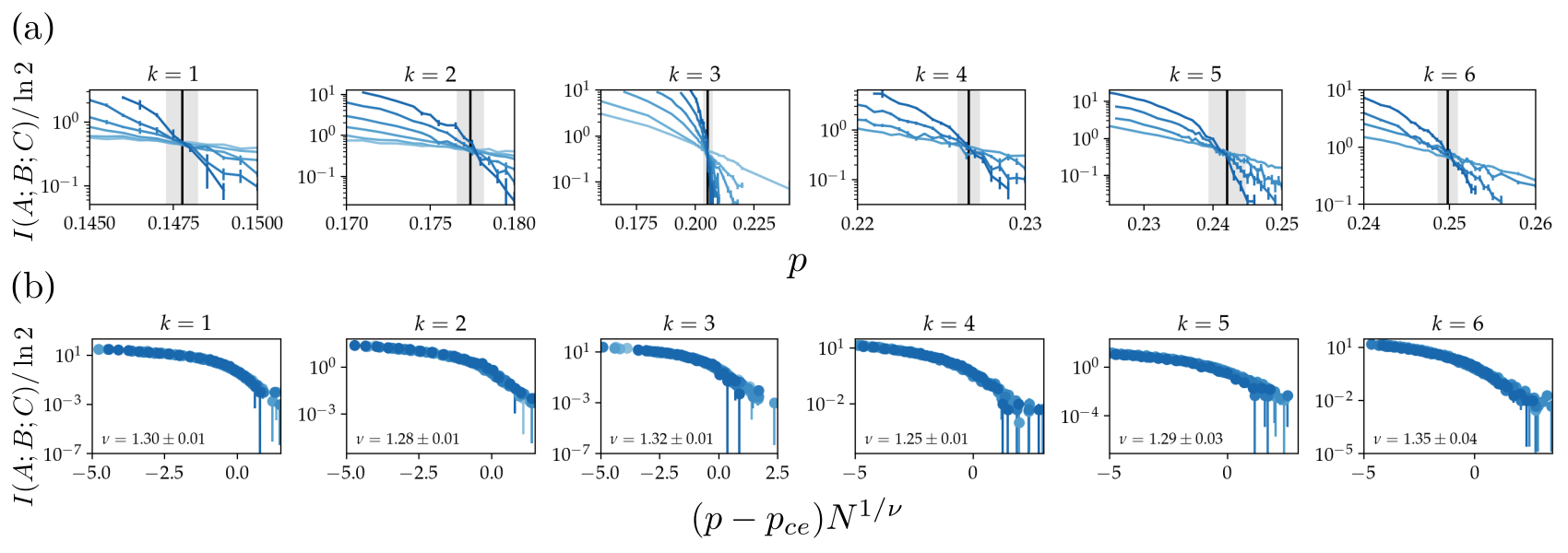}
  \caption{
      (a) The tripartite mutual information $I(A;B;C)$ 
      of initial $z$-polarized state evolved under a deterministic Clifford \pwrk circuit 
      until $t=8N$ for $k=1,2,\dots6$ for the system sizes $N=2^{6},\dots,2^{11}$ (light to dark blue).
      The solid vertical lines are the estimated value of the $p_{ce}$ and the shaded region is the corresponding $1$-sigma errors.
      (b) Collapsed tripartite mutual information for $k=1,2,\dots6$ for the system sizes $N=2^{6},\dots,2^{11}$ (light to dark blue).
      \label{App:FSEntanglement}
  }
\end{figure*}
The finite-size scaling analysis of the entanglement criticality of \pwrk circuit for $k=1,3,5$
was presented in Fig. 3~b of the main text.
Here we show the finite-size scaling and the crossings of the tripartite mutual information $I(A;B;C)$ 
for all the values of $k=1,2,\dots,6$ for system sizes $N=2^{6},\dots,2^{11}$
(the system sizes that are smaller than $2^{k+2}$ are not used in the analysis).

The error bars are estimated by taking the standard error of up to 1000 realizations of the random projective measurement of the 
evolution to $t=8N$.
The critical points and their errors are calculated by computing the average intersection points 
of the curves with 5000 different realizations of the fluctuations added to each of the points 
that are drawn from the appropriate distribution with the standard error as the standard deviation at the point 
(\figref{App:FSEntanglement}~a). 
The critical exponents and their errors are also estimated by collapsing the 5000 different realizations of the fluctuations
added to the data points from the same distribution (\figref{App:FSEntanglement}~b).

\subsection{Finite Size Scaling of the Purification Critical Point for $k=1,\dots,6$}
\label{app:scalingcollapsePur}

\begin{figure*}[t]
  \includegraphics[width=0.95\textwidth]{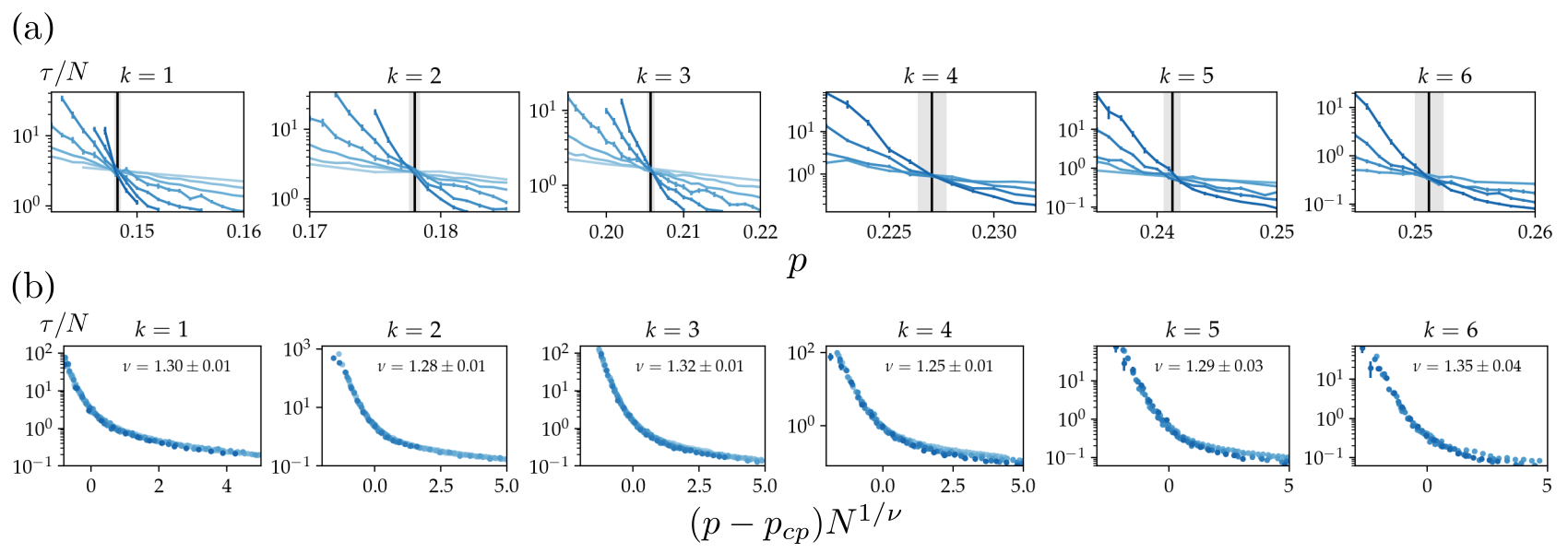}
  \caption{
      (a) The single-qubit purification time for $k=1,2,\dots6$ for the system sizes $N=2^{6},\dots,2^{11}$ (light to dark blue).
      The initial stat is prepared such that there is a qubit entangled to one of the $N$ qubits (system), then the system qubits
      are scrambled with nearest neighbour random Clifford circuit up to $t=4N$.
      The solid vertical lines are the estimated value of the $p_{cp}$ and the shaded region is the corresponding $1$-sigma errors.
      (b) Collapsed tripartite mutual information for $k=1,2,\dots6$ for the system sizes $N=2^{6},2^{11}$ (light to dark blue).
      \label{App:FSPurification}
  }
\end{figure*}

The finite-size scaling analysis of the purification criticality of \pwrk circuit for $k=1,3,5$
was presented in Fig. 4~b of the main text.
Here we show the finite-size scaling and the crossings of the single-qubit purification time 
for all the values of $k=1,2,\dots,6$ for system sizes $N=2^{6},\dots,2^{11}$
(the system sizes that are smaller than $2^{k+2}$ are not used in the analysis). 

The error bars are estimated by taking the standard error of up to 1000 realizations of the random projective measurement of the 
evolution until the state is purified. 
The critical points and their errors are estimated by computing the average intersection points 
of the curves with 5000 different realization of the fluctuations added to each of the points 
that are drawn from the appropriate distribution with standard error as its standard deviation at each points
(\figref{App:FSPurification}~a). 
The critical exponents and their errors are also estimated by collapsing the 5000 different realizations of the fluctuations
added to the data points from the same distribution (\figref{App:FSPurification}~b).

\begin{figure*}[h!]
  \includegraphics[width=0.95\textwidth]{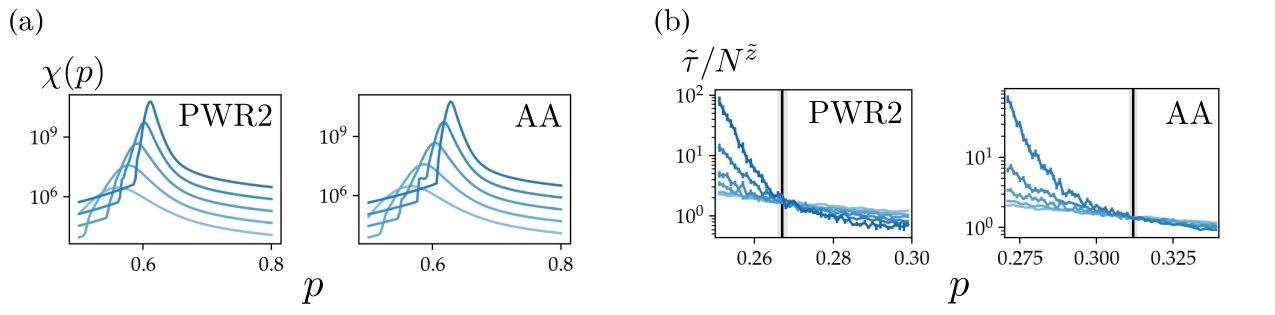}
  \caption{
      (a) Susceptibility, $\chi(p)$, of the complete \pwrk (left) and AA (right) percolation networks
      for the system sizes $N=2^{6},2^{7},\dots,2^{10}$.
      Although there exists sharp peaks, the positions of the peaks depends significantly on the system size. 
      The size of the circuits in this calculations is taken to be $N \times T$.
      The observables of the bond percolation of the networks
      are calculated with Newman-Ziff algorithm by taking the microcanonical ensemble of over 4000 trajectories. 
      (b) $\tilde{\tau}/N^{\tilde{z}}$ as a function of $p$
      without the scaling collapse for complete \pwrk (left) and AA (right) circuits 
      for the system sizes $N=2^{6},2^{7},\dots,2^{10}$ with $\tilde{z}=0.21\pm0.01$ (\pwr)
      and $\tilde{z}=0.170\pm0.006$ (AA).
      The critical point determined from the scaling collapse is marked by the black line 
      and the 1-sigma error is marked by the grey shade.
      The critical points and their error are $\tilde{p}_{c}=0.267\pm0.001$ and $\tilde{p}_{c}=0.312\pm0.001$ respectively.
      \label{App:FSPWR2AA}
  }
\end{figure*}

\subsection{Finite Size Scaling of the Full \pwr and AA circuits}
\label{app:FSFullandAA}

The finite-size scaling of the MIPTs of the complete \pwrk circuit is difficult due to the long-range interactions
which give rise to strong boundary effects and the loss of locality.
In the percolation transition, this appears as the strong system size dependence of the 
$p_{\mathrm{peak},\mathrm{Haar}}$ as shown in Fig. 6~a of the main text.
The positions of the peaks are determined from the $\chi(p)$ plotted in \figref{App:FSPWR2AA}~a.

The loss of locality makes the determination of the entanglement transition impossible.
Due to the coupling of distance $N/2$, even in the regime near $p=1$,
the entanglement entropy of the region of size $A<N/2$ is almost guaranteed to be $\sim A(1-p)$ 
at the end of the application of $2\log_2 N-1$ layers of the gates.
Therefore we do not argue on the existence of the entanglement transition in the complete \pwrk and AA circuits.

The percolation transition suffers from the similar boundary problem.
However, the transition does not ask the geometry of the circuit as it only asks the entropy of the reference qubits that 
are entangled to the system at $t=0$.
The problem in the determination of the critical properties of the purification transition is 
the number of layers of the even (or odd) blocks which depends on the system size.
In this paper, we proposed the potential workaround, which is normalizing the time 
$t$ by the number of layers in the even (or odd) blocks by defining $\tilde{t}=t/\log_2 N$.
This lead to the empirical scaling law of the form in Eq. (7) of the main text.

As shown in Fig. 6~d, the single-qubit purification time $\tau(p)$ 
collapsed surprisingly well with Eq. (7) of the main text, with 
$\tilde{p}_c=0.27\pm0.01$,$\tilde{\nu}_c=2.08\pm0.08$, and $\tilde{z}_c=0.21\pm0.01$ (\pwr) and 
$\tilde{p}_c=0.31\pm0.01$,$\tilde{\nu}_c=2.29\pm0.05$, and $\tilde{z}_c=0.170\pm0.006$ (AA).
For this simulation, $\tilde{\tau}(p)$ is estimated by taking the average over up to 
500 realizations of the random projective measurements.
The errors in the critical exponents and critical points are estimated by performing the scaling collapse on the 
5000 different realizations of the data set with random fluctuations.
The random fluctuations are taken from the appropriate distribution with standard deviation of the standard error at each point.

\subsection{The Code Distances at Logarithmic Time Scale}
\label{app:subcodedistancelogscale}

\begin{figure*}[htb]
   \includegraphics{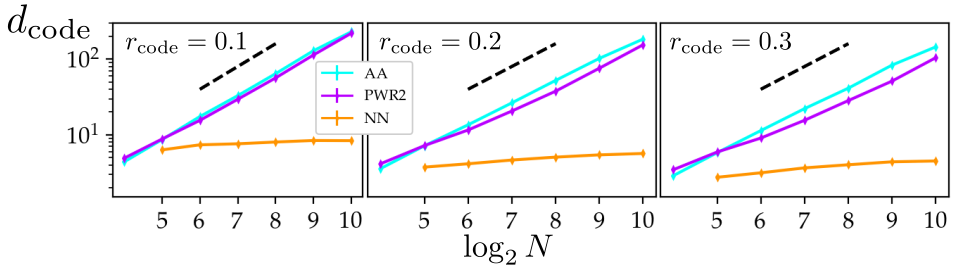}
   \caption{
      System size scaling of the contiguous code distances $d_{\mathrm{code}}$ at the circuit depth $T=8\log_2 N$
      for code rates $r_{\mathrm{code}}=0.1,0.2,0.3$.
      $d_{\mathrm{code}}\propto N^{\beta}$ with $\beta\sim 0.9$ for AA and PWR2 for all the code rates. 
      NN on the other hand, the polynomial dependency is not observed, in this very early times.
      The black dotted lines are there for the guide for linearly scaling code distance ($d_{\mathrm{code}} \propto N$).
   \label{fig:CD_logN}}
\end{figure*}
In the previous subsections, the critical exponents and code distances were determined on the output state after the 
evolution for the total circuit depth, $T$, which is linearly increasing with the system size $N$. 
In this subsection, we look at how the contiguous code distance scales with $T$ which increases 
logarithmically with the system size ($T=8\log_2 N$).

Shown in \figref{fig:CD_logN} is the contiguous code distance at $T=8\log_2N$.
In the case of \pwr and All-to-All circuits, even at this early time scales, 
the contiguous code distance $d_{\mathrm{code}}$, increases almost linearly with the system size; 
while in the case of NN circuit, such polynomial dependency is not observed in this very early times. 

\section{The Gap Between Entanglement ($p_{ce}$) and Purification ($p_{cp}$) Critical Points of \pwrk}
\label{app:GapPhase}

\begin{figure}[h!]
    \includegraphics[width=0.5\textwidth]{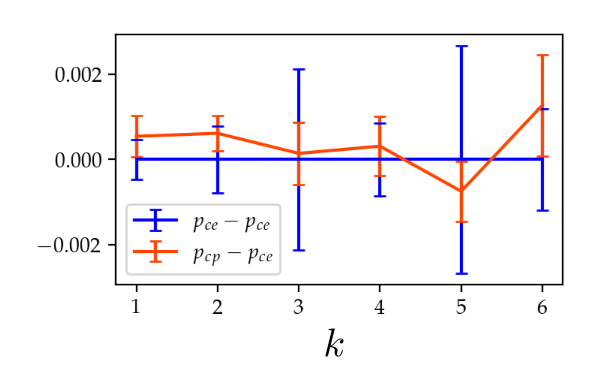}
    \caption{
        The difference between entanglement and purification critical points of \pwrk circuit 
        are plotted (red) for $k=1,\dots,6$.
        The blue line shows the error of entanglement critical points centered at the expected value.
        There is no statistically significant deviation observed between the entanglement and purification critical points. 
        \label{fig:AppGap}
    }
\end{figure}
Whether there exists a gap between the entanglement and purification critical points is a topic of an ongoing debate.
Gullans and Huse \cite{gullans2020dynamical} show that in 1+1D, they occur at the same point in general.
Here we show that such a gap is not observed in the \pwrk circuits for the values of $k=1,\dots,6$.
No statistically significant deviations between the entanglement and the purification critical points are observed
(\figref{fig:AppGap}).

\section{Quantum Error-Correcting Code Properties}
\label{app:qecc}

\subsection{Code Distance for Stabilizer Circuits}
\label{app:subcodedistance}

Here we review an argument of Li and Fisher that connects the code distance $\dcode$ of a stabilizer circuit to the mutual information $I(A,R)$ between a subregion $A \subset Q$ of the output qubits and a maximally-entangled reference $R$ \cite{li2020statistical}. In this work we are primarily interested in \emph{contiguous} regions $A$, but the arguments in this Appendix hold equally well for arbitrary, possibly disconnected regions. For any bipartition $A \cup \overline{A} = Q$ of the system we define the quotient group
\begin{equation}
    \mathcal{L}_A \equiv \frac{\{ g \in \centr | \projab(g) \in \projab(\stabs) \} }{\stabs}.
    \label{eq:lagroupdefn}
\end{equation}
Here $\projab$ is the group homomorphism from $\mathcal{P}(Q)$ to $\mathcal{P}(\overline{A})$ defined by
\begin{equation}
    \projab: g_A \otimes g_{\overline{A}} \rightarrow g_{\overline{A}}
\end{equation}
which projects any Pauli string $g = g_A \otimes g_{\overline{A}}$ down to its support on the region $\overline{A}$. The group $\mathcal{L}_A$ is the group of undetectable errors that are localizable on $A$. In other words, any $g$ in this group can be localized to $A$ simply by multiplying by a suitable choice of $g' \in \stabs$. The order $\magn{\mathcal{L}_A} = 2^{\ell_A}$ of this group gives the number of independent undetectable errors that can be localized to the region $A$. We must have $\ell_A = 0$ in order for a stabilizer code to successfully protect quantum information from errors acting on $A$.

In the following we compute the order $\magn{\mathcal{L}_A}$ of the group $\mathcal{L}_A$ and show that $\ell_A \equiv \log_2 \magn{\mathcal{L}_A}$ is proportional to the mutual information $I(A,R)$. Starting from the definition \eqref{eq:lagroupdefn}, we find that the order of the group can be written
\begin{align}
    \magn{\mathcal{L}_A} = \frac{\magn{\centr} \cdot \magn{\projab(\stabs)}}{\magn{\stabs} \cdot \magn{\projab(\centr)}}
    \label{eq:laorder}
\end{align}
(see Appendix A of \cite{li2020statistical} for a rigorous group-theoretic derivation).
Our goal is to convert the quantities on the RHS into Renyi entropies, allowing us to relate $\ell_A$ to $I(A,R)$.

As a preliminary step, we first show how to compute the Renyi entropy $S^{(2)}(\rho_A(\stabs))$ of a reduced density matrix $\rho_A(\stabs)$ for a subregion $A \subset Q$ of the full system. The Renyi entropy of such a region $A$ is given by:
\begin{align}
    \rho_A(\stabs) &= \trover{\overline{A}}{\rho_Q(\stabs)} \nonumber \\
    &= 2^{-\magn{Q}} \sum_{g \in \stabs}\trover{\overline{A}}{g} \nonumber \\
    &= 2^{-\magn{A}} \sum_{g \in \stabs \cap \mathrm{Ker} \ \projab} g
\end{align}
where in the third line we used the fact that $\trover{\overline{A}}{g} = 0$ except when $\projab(g) = \mathbb{I}_{\overline{A}}$ (i.e. $g \in \mathrm{Ker} \ \projab$). Introducing the group $\mathcal{S}_A \equiv \stabs \cap \mathrm{Ker} \ \projab \cong \stabs / \projab(\stabs)$ consisting of all stabilizer operators that act trivially outside the subregion $A$, we can simply write the reduced density matrix as 
\begin{equation}
    \rho_A(\stabs) = 2^{-\magn{A}} \sum_{g \in \stabs_A} g
\end{equation}
which is a natural generalization of \eqref{eq:codestate}. Finally, it is easy to show that the Renyi entropy of the reduced state $\rho_A(\stabs)$ is given by
\begin{align}
    (\ln 2)^{-1} S^{(2)}(\rho_A(\stabs)) &= \magn{A} - \log_2 \magn{\stabs_A} \nonumber \\
    &= \magn{A} - \log_2 \magn{\stabs} + \log_2 \magn{\projab(\stabs)}
    \label{eq:renyientropy}
\end{align}
where we have used $\magn{\stabs_A} = \magn{\stabs} / \magn{\projab(\stabs)}$.

Finally, combining \eqref{eq:laorder} and \eqref{eq:renyientropy} we find:
\begin{widetext}
\begin{align*}
    (\ln 2)^{-1} I(A,R) &= (\ln 2)^{-1} \left( S_R + S_A - S_{AR} \right) \nonumber \\
    &= \magn{R} + (\ln 2)^{-1} S_A - (\ln 2)^{-1} S_{AR} + (\log_2 \magn{\mathcal{C}(\stabs)} -  \log_2 \magn{\mathcal{C}(\stabs)}) + (\log_2 \magn{\stabs} - \log_2 \magn{\stabs}) \nonumber \\
    &= \log_2 \magn{\mathcal{C}(\stabs)} - \log_2 \magn{\stabs} + \left( (\ln 2)^{-1} S_{A} - \magn{A} + \log_2 \magn{\stabs} \right) - \left( (\ln 2)^{-1} S_{AR} - \magn{AR} + \log_2 \magn{\mathcal{C}(\stabs)} \right) \nonumber \\
    &= \log_2 \magn{\mathcal{C}(\stabs)} - \log_2 \magn{\stabs} + \log_2 \magn{\projab(\stabs)} - \log_2 \magn{\projab(\mathcal{C}(\stabs))} \nonumber \\
    &= \log_2 \magn{\mathcal{L}_A} = \ell_A
\end{align*}
\end{widetext}
where we have used the shorthand $S_A \equiv S^{(2)}(\rho_A(\stabs))$ to shorten notation, and in going from the third to fourth line we have used the identity \eqref{eq:renyientropy} and its corresponding generalization for the Renyi entropy $S_{AR}$ of the combined regions $A,R$ \cite{li2020statistical}.

\subsection{Quantum Hamming Bound}
\label{app:subhammingbound}

Here we review the quantum Hamming bound, which places a fundamental bound on achievable code rates and code distances for correctable nondegenerate QECCs. Suppose we have a QECC with rate $\rcode = K/N$ that can correct all errors of weight less than or equal to $w$. (Note that $K \neq k$, where $k$ is the nonlocality parameter from the main text.) There are a total of $\sum_{i=0}^w 3^i \binom{N}{i}$ such errors (Pauli strings). For nondegenerate codes, we must have $\bra{\psi} E E' \ket{\psi'} = 0$ for all errors $E,E'$ and all code states $\ket{\psi},\ket{\psi'}$. That is, every code state $\ket{\psi}$ and all of its erroneous versions must be orthogonal to every other code state and all of its erroneous versions \cite{steane2006tutorial}. The only way all of these orthogonal states fit into the same $2^N$-dimensional Hilbert space is when
\begin{equation}
    2^K \sum_{i=0}^w 3^i \binom{N}{i} \leq 2^N
\end{equation}
For large $N$ and fixed $K/N,w/N$ this is equivalent to
\begin{equation}
    \frac{K}{N} \leq 1 - \frac{w}{N} \log_2 3 - H(w/N)
\end{equation}
where $H(x) = -x \log x - (1-x) \log (1-x)$ is the classical Shannon entropy.

One can derive a similar bound for the case where errors only occur contiguously. In this case, there are a total of $\sum_{i=0}^w 3^i (N+1-i)$ contiguous errors of weight less than or equal to $w$. For large $N$ and fixed $K/N,w/N$ this yields a bound
\begin{equation}
    \frac{K}{N} \leq 1 - \frac{w}{N} \log_2 3 - \frac{1}{N} \log_2 (N-w)
\end{equation}
where the final term vanishes in the strict limit $N \rightarrow \infty$, but is still relevant for large but finite size systems.

\subsection{Practical Error Correction: Encoding and Decoding}
\label{app:subpracticalerrorcorrection}

The non-unitary circuits we study in this work have desirable quantum error-correcting code properties -- to what extent can we actually harness these circuits to perform useful error-correction in real-world applications? Two primary complications arise. First, the measurement outcomes obtained in the laboratory are random, meaning that the non-unitary circuit $V$ is different on every experimental run; so how can we reliably use the encoding circuit $V$ for error correction if it changes every time we run the experiment? Second, once the information is encoded, how does one perform error correction in practice (decoding)? Here we explicitly demonstrate how to use hybrid Clifford circuits to perform encoding and decoding of quantum information in real-world applications, despite the fact that the encoding circuit $V$ changes on each experimental run. This strategy crucially relies on the fact that the circuit $V$ consists entirely of Clifford operators so that it can be classically simulated, and does not generalize to circuits containing non-Clifford elements.
Nevertheless, these Clifford-only encoding circuits are able to store arbitrary coherent quantum information, which need not be limited to stabilizer states. This is similar to the situation in conventional stabilizer QECCs which are capable of storing arbitrary coherent quantum information, despite the fact that the stabilizer matrix (parity-check matrix) $H$ consists only of Clifford operators. In fact, one can view our non-unitary circuits $V$ as preparing a random stabilizer code, whose stabilizer matrix $H$ can be obtained via classical post-processing of the measurement results as we explain below.

\begin{figure}[h!]
    \centering
    \includegraphics[width=0.5\textwidth]{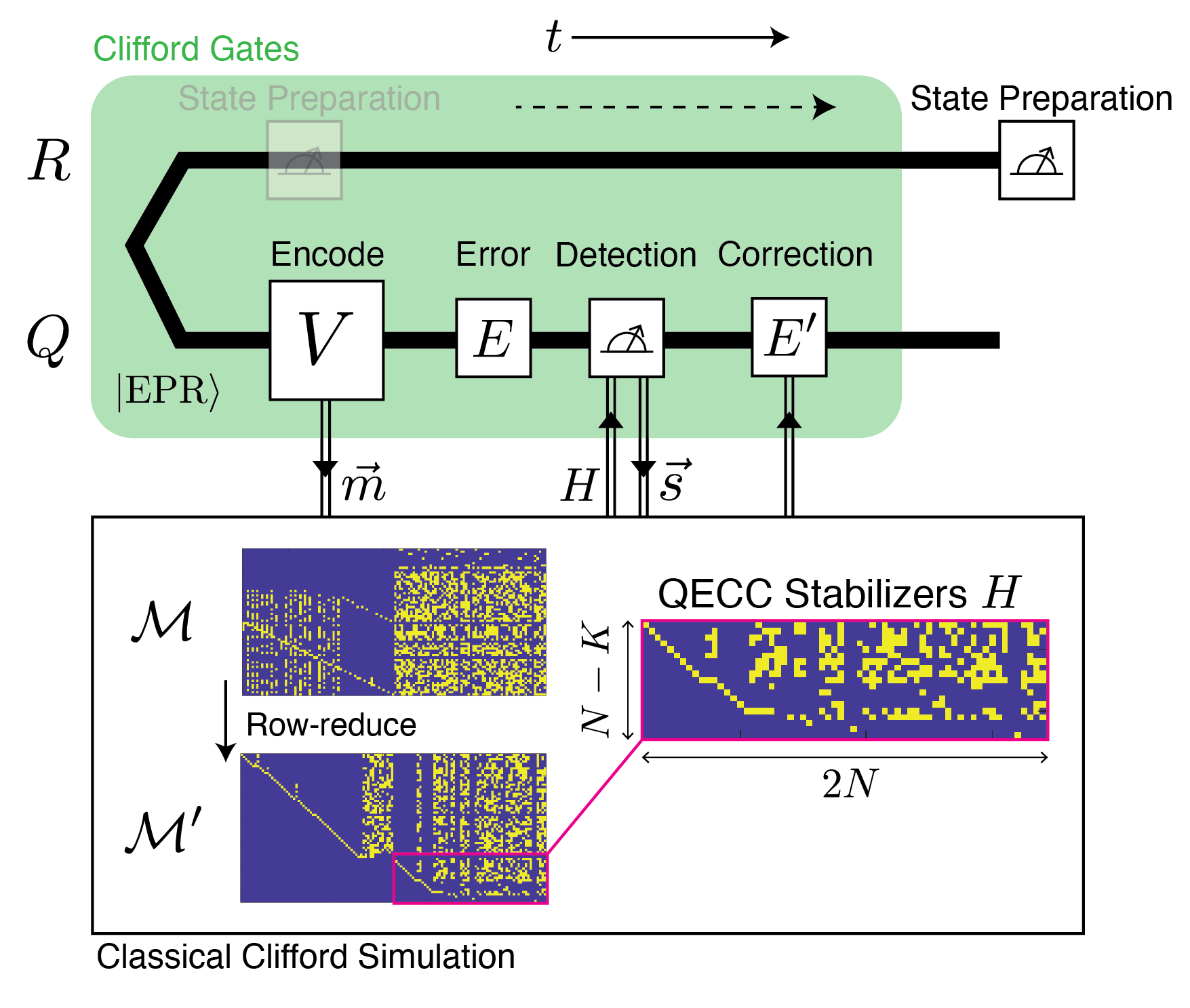}
    \caption{Encoding and decoding procedure for harnessing non-unitary Clifford circuits $V$ for practical quantum error correction. Operations performed inside the green box (top) consist entirely of Clifford operations and can therefore be simulated on a classical computer (bottom). Arbitrary quantum states can be prepared by projectively measuring the reference qubits $R$. The principle of deferred measurement allows us to consider this state preparation to occur outside of the green Clifford-only box.}
    \label{fig:miptdecodecircuit}
\end{figure}

\begin{figure*}
    \centering
    \includegraphics[width=\textwidth]{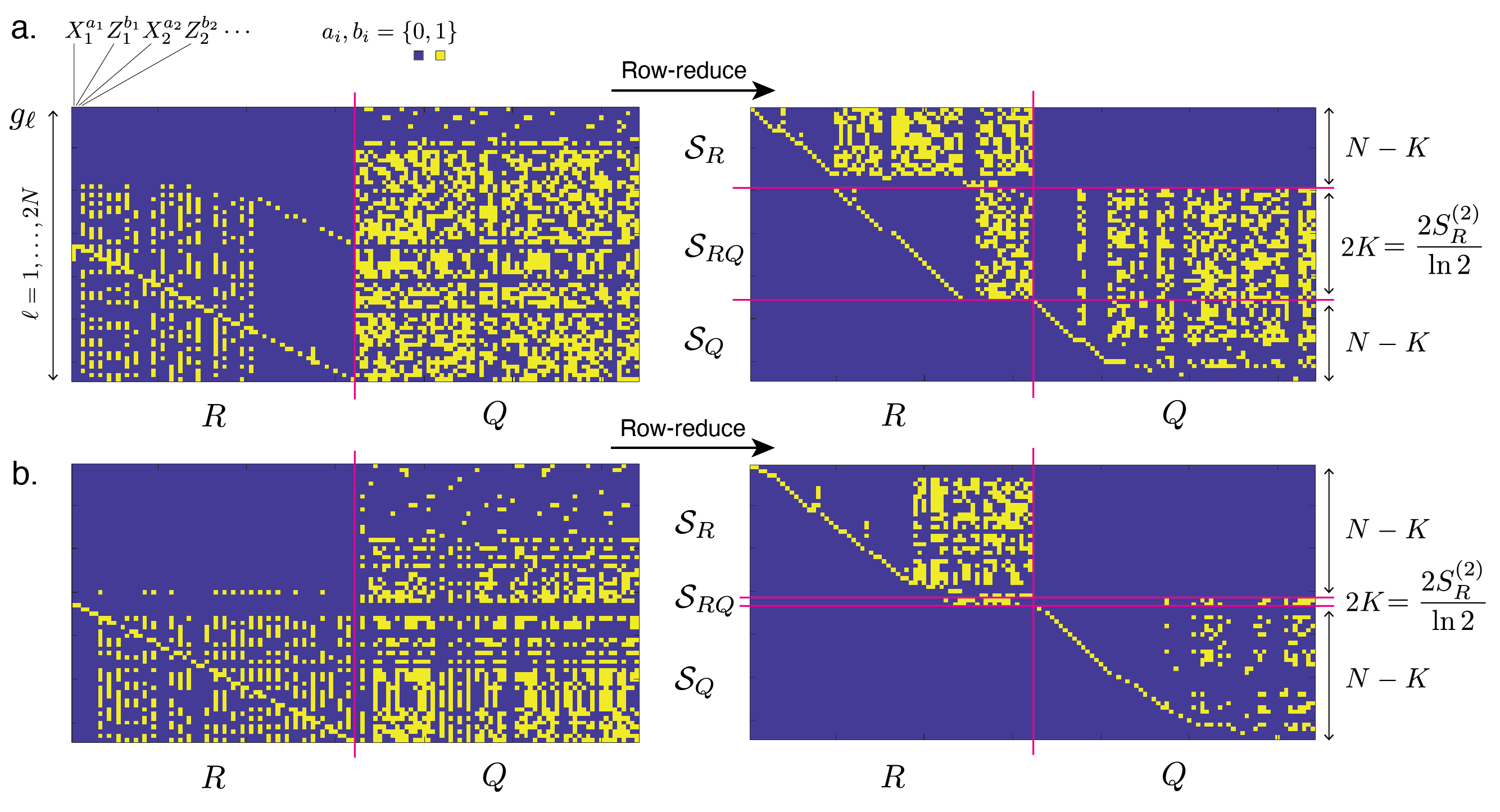}
    \caption{Stabilizer matrices for quantum error-correcting code states generated by MIPT dynamics in the `complete' PWR2 circuit on $N = 32$ qubits with measurement rate $p = 0.05$ (a) and $p = 0.15$ (b). Each row $\ell = 1,\ldots,2N$ (left) represents a stabilizer $g_{\ell}$, whose operator content is split between the reference $R$ and the system $Q$. Binary row-reduction yields an equivalent representation (right) in which the stabilizer group has been explicitly decomposed into components $\mathcal{S} = \mathcal{S}_R \times \mathcal{S}_{RQ} \times \mathcal{S}_Q$, where $\mathcal{S}_R,\mathcal{S}_Q$ comprise all stabilizers that can be localized to $R$,$Q$, respectively, and $\mathcal{S}_{RQ}$ comprise the remaining stabilizers which necessarily feature nontrivial operator content on both $R$ and $Q$. The $(N-K) \times 2N$ sub-matrix in the lower right-hand corner of the row-reduced stabilizer matrix serves as the stabilizer matrix (parity-check matrix) $H$ of the resulting stabilizer QECC, while the middle $2 K$ rows capture entanglement between $Q,R$ and can be used to store quantum information in the system, where the rate of the code is given by $\rcode = K/N$.}
    \label{fig:miptdecode}
\end{figure*}

The encoding and decoding process is shown in Fig. \ref{fig:miptdecodecircuit}. In the first step, the system $Q$ and reference $R$ are maximally entangled via a collection of EPR pairs, and the non-unitary circuit $V$ is applied to the system $Q$. The results of all projective measurements in $V$ are stored in a classical computer as a measurement record $\vec{m}$. Using this measurement record, along with our knowledge of the gate sequence in the circuit $V$, we may classically simulate the resulting dynamics of the $2N$-qubit system $QR$ \cite{aaronson2004improved}. Thus at the end of each experimental run we obtain a stabilizer group $\stabs$ of dimension $\magn{\stabs} = 2^{2N}$ corresponding to a pure state $\ket{\chi} = \id^R \otimes V^Q \ket{\mathrm{EPR}}$ of $\magn{Q} + \magn{R} = 2N$ qubits, where $\id^R$ is the identity acting on $R$ and $V^Q$ is the non-unitary circuit $V$ acting on $Q$. The stabilizer group $\stabs$ is represented by a $2N \times 4N$ binary stabilizer matrix $\mathcal{M}$ where each row $\ell = 1,\ldots,2N$ corresponds to a stabilizer generator $g_{\ell}$ as shown on the left side of Fig. \ref{fig:miptdecode}. By performing row-reduction on $\mathcal{M}$ -- equivalent to a basis transformation $g_{\ell} \rightarrow g_{\ell}'$ among the stabilizer generators -- we can bring the matrix to reduced row-echelon form $\mathcal{M}'$. By construction, the bottom $N-K$ rows of the row-reduced matrix $\mathcal{M}'$ generate the subgroup $\mathcal{S}_Q = \stabs / \proj_{R} (\stabs)$, the subgroup of stabilizers that can be completely localized to the system $Q$. This group defines the stabilizer group $\mathcal{H} \equiv \mathcal{S}_Q$ of our quantum error-correcting code: in particular, the $(N-K) \times 2N$ sub-matrix in the lower-right corner of $\mathcal{M}'$ serves as the stabilizer matrix (parity-check matrix) $H$ of the QECC. This code can correct any error $E$ that fails to commute with one or more of the stabilizers in $H$ by using standard error-correction procedures for stabilizer codes \cite{NielsenChuang2010,steane2006tutorial}.

So far, this procedure tells us how to detect and correct errors, but it does not tell us how to store a particular quantum state $\ket{\psi}$ in the system. To do this we make use of the initial entanglement between $Q$ and $R$. Some of this entanglement is destroyed by the projective measurements in $V$, but in the volume-law phase $K$ bits of entanglement remain between $Q,R$ that can be harnessed to encode any $K$-qubit state into the QECC. Further row-reduction operations on $\mathcal{M}$ can bring the $QR$ stabilizer matrix into the form shown on the right side of Fig. \ref{fig:miptdecode}, which gives a decomposition of the complete stabilizer group into components $\mathcal{S} = \mathcal{S}_R \times \mathcal{S}_{RQ} \times \mathcal{S}_Q$, where $\mathcal{S}_{RQ} = \mathcal{S} / (\mathcal{S}_R \times \mathcal{S}_Q)$. As discussed above, the subgroup $\mathcal{H} = \mathcal{S}_Q$ serves as a stabilizer code for the system $Q$ (similarly, the subgroup $\mathcal{S}_R$ could serve as a stabilizer code for the reference $R$ if the roles of $Q,R$ were reversed). The remaining subgroup $\mathcal{S}_{RQ}$ is generated by $2 K$ stabilizer operators with support on both $Q$ and $R$. This subgroup describes the remaining $K$ bits of entanglement that survive between $Q$ and $R$ despite the measurements performed in the non-unitary dynamics $V^Q$.

We now describe how to leverage this remaining entanglement to store a desired $K$-qubit quantum state $\ket{\psi}$ in the system $Q$ by preparing a time-reversed version of the state $\ket{\psi'}$ on the reference $R$. Further row-reduction operations within the subspace $\mathcal{S}_{RQ}$ can be used to transform the $2 K$ stabilizers into a standard generating set $\mathcal{S}_{RQ} = \{g_{r}\}$ \cite{aaronson2004improved} of the form
\begin{align}
    g_{r} = \begin{cases}
        \overline{X}_i^R \otimes  \overline{X}_i^Q \quad r = 2i-1 \\
    \overline{Z}_i^R \otimes \overline{Z}_i^Q \quad r = 2i
    \end{cases}
    \label{eq:klogicalgenerators}
\end{align}
for $i = 1,\ldots,K$, where $\overline{X}_i^R,\overline{Z}_i^R$ are many-body Pauli strings with support only on $R$ and $\overline{X}_i^Q,\overline{Z}_i^Q$ are many-body Pauli strings with support only on $Q$. In this standard form, these operators all mutually commute $[\overline{X}_i,\overline{X}_j] = [\overline{X}_i,\overline{Z}_j] = [\overline{Z}_i,\overline{Z}_j] = 0$ when $i \neq j$, but when $i = j$ we have $\{\overline{X}_i^R,\overline{Z}_i^R\} = \{\overline{X}_i^Q,\overline{Z}_i^Q\} = 0$. As suggested by the notation, the operators $\overline{X}_i^Q,\overline{Z}_i^Q$ thus define a set of logical bit-flip and phase-flip (or `shift' and `clock') operators for $K$ logical qubits on the system $Q$. Similarly, the operators $\overline{X}_i^R,\overline{Z}_i^R$ represent logical bit-flip and phase-flip operators for $K$ logical qubits on the reference $R$. The form of the generators $g_r$ in Eq. \eqref{eq:klogicalgenerators} guarantees that each logical qubit $i$ in $Q$ is maximally entangled with a corresponding logical qubit $i$ in $R$.

This entanglement between logical qubits in $Q$ and $R$ immediately allows us to store a $K$-qubit quantum state $\ket{\psi} \bra{\psi}_Q$ in the system $Q$ by preparing the reference logical qubits in a related state $\ket{\psi'} \bra{\psi'}_R$. As a warmup, consider preparing a single-qubit state $\ket{\psi} \bra{\psi}_{\ell} = (\id - \vec{n} \cdot \vec{\sigma}_{\ell})/2$ on the left half of an EPR pair $\ket{\mathrm{EPR}}_{\ell r} = (\ket{01} - \ket{10})_{\ell r} / \sqrt{2}$ by projecting the right half onto the state $\ket{\psi'} \bra{\psi'}_r = (\id - \vec{n}^T \cdot \vec{\sigma}_r)/2$. Here $\vec{n} = (n_x,n_y,n_z)$ and $\vec{n}^T = (n_x,-n_y,n_z)$ are the unit-length Bloch sphere vectors for the left and right qubits, respectively, where the additional minus sign accounts for the necessary time-reversal of the qubit state required in passing the state from the left to the right side of the EPR pair.
We can perform a similar procedure in the many-body case, making similar use of the $K$ qubits of entanglement shared between $Q,R$. For simplicity, we consider encoding a \emph{separable} $K$-qubit state $\ket{\psi} \bra{\psi} = \frac{1}{2^K} \prod_{i=1}^K \left( \id - \vec{n}_i \cdot \vec{\sigma}_i \right)$, where the unit vectors $\vec{n}_i = (n_i^x, n_i^y, n_i^z)$ are the Bloch sphere vectors of the $K$ qubits and $\vec{\sigma}_i = (\sigma_i^x,\sigma_i^y,\sigma_i^z)$ are the standard Pauli matrices for each qubit $i$. To store this state in the system $Q$, we prepare the reference $R$ in a time-reversed state defined by the rank-1 projector
\begin{equation}
    \ket{\psi'} \bra{\psi'}_R = \mathbb{P}^R_{\vec{n}_i^T} = \frac{1}{2^K} \prod_{i=1}^K \left( \id - \vec{n}_i^T \cdot \vec{\sigma}_i^R \right)
\end{equation}
where $\vec{\sigma}_i^R = (\overline{X}_i^R, i \overline{X}_i^R \overline{Z}_i^R, \overline{Z}_i^R )$ and $\vec{n}_i^T = (n_i^x, -n_i^y, n_i^z)$. The additional minus sign in $n_i^y$ accounts for time-reversal associated with the EPR pairs shared between $Q,R$ as mentioned earlier. This prepares the quantum state
\begin{equation}
    \ket{\overline{\psi}} = \mathbb{P}_{\vec{n}_i^T}^R \ket{\chi},
\end{equation}
which we claim is a logical encoding of the separable $K$-qubit state $\ket{\psi}$ defined by the vectors $\vec{n}_i$ in the system $Q$. This information is protected from errors by the stabilizer QECC discussed above.

To prove that the state $\ket{\overline{\psi}}$ does indeed store this quantum information in the system $Q$, we can measure the projector
\begin{equation}
    \ket{\psi} \bra{\psi}_Q = \mathbb{P}^Q_{\vec{n}_i} = \frac{1}{2^K} \prod_{i=1}^K \left( \id - \vec{n}_i \cdot \vec{\sigma}_i^Q \right)
\end{equation}
which acts only on the system $Q$, where $\vec{\sigma}_i^Q = (\overline{X}_i^Q, i \overline{X}_i^Q \overline{Z}_i^Q, \overline{Z}_i^Q )$ and $\vec{n}_i = (n_i^x,n_i^y,n_i^z)$. This yields
\begin{align}
    \mathbb{P}^Q_{\vec{n}_i} \ket{\overline{\psi}} &= \mathbb{P}^Q_{\vec{n}_i} \mathbb{P}^R_{\vec{n}_i^T} \ket{\chi} \nonumber \\
    &= \left(\mathbb{P}^R_{\vec{n}_i^T}\right)^2 \ket{\chi} \nonumber \\
    &= \mathbb{P}^R_{\vec{n}_i^T} \ket{\chi} = \ket{\overline{\psi}}
\end{align}
and thus $\ket{\overline{\psi}}$ is an eigenstate of the projector $\mathbb{P}_{\vec{n}_i}^Q$. In the second line above we have converted the system projector $\mathbb{P}^Q_{\vec{n}_i}$ into a reference projector $\mathbb{P}^R_{\vec{n}_i^T}$ by `pulling the operator through' the EPR pairs in $\ket{\chi}$ via the identities
\begin{align}
    \overline{X}_i^Q \ket{\chi} &= \overline{X}_i^R \ket{\chi} \nonumber \\
    \overline{Z}_i^Q \ket{\chi} &= \overline{Z}_i^R \ket{\chi}
\end{align}
These identities are immediate consequences of the stabilizer condition $g_r \ket{\chi} = + \ket{\chi}$ in the subspace $\mathcal{S}_{RQ} = \{g_r\}$ along with the fact that the logical bit-flip and phase-flip operators $\overline{X}_i^R,\overline{Z}_i^R$ square to the identity.
The fact that $\ket{\overline{\psi}}$ is an eigenstate of both projectors $\mathbb{P}_{\vec{n}_i}^Q, \mathbb{P}_{\vec{n}_i^T}^R$ immediately implies that
\begin{equation}
    \ket{\overline{\psi}} \bra{\overline{\psi}} = \mathbb{P}^Q_{\vec{n}_i} \otimes \mathbb{P}^R_{\vec{n}_i^T} = \ket{\psi} \bra{\psi}_Q \otimes \ket{\psi'} \bra{\psi'}_R
\end{equation}
identically. We have thus prepared a quantum state $\ket{\psi} \bra{\psi}_Q$ on the system $Q$ by projecting the entangled reference $R$ onto a time-reversed state $\ket{\psi'} \bra{\psi'}_R$.
These arguments readily generalize to arbitrary nonseparable $K$-qubit states, meaning that we can store arbitrary quantum information in this QECC by preparing an appropriate time-reversed state on the entangled reference qubits.

\clearpage

\end{document}